\documentclass[10pt,journal,compsoc]{IEEEtran}               

\usepackage{mathptmx}
\usepackage{graphicx}
\usepackage{times}
\usepackage{amssymb}
\usepackage{cite} 
\usepackage{amsfonts}
\usepackage{algorithm2e}
\usepackage{enumitem} 
\usepackage{titlesec}
\usepackage{xspace}
\usepackage{float}
\usepackage{soul}
\usepackage[dvipsnames]{xcolor}

\makeatletter
\newcommand{\nosemic}{\renewcommand{\@endalgocfline}{\relax}}
\newcommand{\dosemic}{\renewcommand{\@endalgocfline}{\algocf@endline}}
\makeatother

\SetKwInput{KwTitle}{Procedure}
\SetKwInput{KwData}{Input}
\SetKwInput{KwResult}{Output}
\SetKwFor{for}{for (}{) $\lbrace$}{$\rbrace$}

\newcommand{\myparagraph}[1]{\vspace{1mm} \noindent \textbf{#1}}
\newcommand{\ie}{\textit{i.e.},\xspace}
\newcommand{\gmsc}{g\textsc{msc}\xspace}
\newcommand{\ttk}{\textsc{ttk}\xspace}

\usepackage[bookmarks,backref=page,linkcolor=black]{hyperref} 
\hypersetup{
  pdfauthor = {},
  pdftitle = {},
  pdfsubject = {},
  pdfkeywords = {},
  colorlinks=true,
  linkcolor= black,
  citecolor= black,
  pageanchor=true,
  urlcolor = black,
  plainpages = false,
  linktocpage
}

\begin{document}

\title{A GPU Parallel Algorithm for Computing Morse-Smale Complexes}



\author{Varshini~Subhash,
        Karran~Pandey,
        and~Vijay~Natarajan~\IEEEmembership{Member,~IEEE}
\IEEEcompsocitemizethanks{\IEEEcompsocthanksitem Varshini Subhash, Karran Pandey, and Vijay Natarajan are with the Department of Computer Science and Automation, Indian Institute of Science, Bangalore, 560012.\protect\\
E-mail: varshini96@gmail.com, karran13@gmail.com, vijayn@iisc.ac.in.}
\thanks{Manuscript received August X, 2021; revised XXXX, 2021.}}

\markboth{IEEE Transactions on Visualization and Computer Graphics,~Vol.~X, No.~X, Month~Year}%
{Subhash \MakeLowercase{\textit{et al.}}: A GPU Parallel Algorithm for Computing Morse-Smale Complexes}


\IEEEtitleabstractindextext{%
\begin{abstract}
The Morse-Smale complex is a well studied topological structure that represents the gradient flow behavior between critical points of a scalar function. It supports multi-scale topological analysis and visualization of feature-rich scientific data. Several parallel algorithms have been proposed towards the fast computation of the 3D Morse-Smale complex. Its computation continues to pose significant algorithmic challenges. In particular, the non-trivial structure of the connections between the saddle critical points are not amenable to parallel computation. This paper describes a fine grained parallel algorithm for computing the Morse-Smale complex and a GPU implementation (\gmsc). The algorithm first determines the saddle-saddle reachability via a transformation into a sequence of vector operations, and next computes the paths between saddles by transforming it into a sequence of matrix operations. Computational experiments show that the method achieves up to 8.6$\times$ speedup over pyms3d and 6$\times$ speedup over TTK, the current shared memory implementations. The paper also presents a comprehensive experimental analysis of different steps of the algorithm and reports on their contribution towards runtime performance. Finally, it introduces a CPU based data parallel algorithm for simplifying the Morse-Smale complex via iterative critical point pair cancellation.
\end{abstract}

}








\maketitle

\IEEEpeerreviewmaketitle
\IEEEraisesectionheading{\section{Introduction}\label{sec:introduction}}


The Morse-Smale (MS) complex~\cite{EHZ03, EHNP03} is a topological structure that provides an abstract representation of the gradient flow of a scalar function. It represents a decomposition of the domain of the scalar field into regions with uniform gradient flow behavior. Applications to feature-driven analysis and visualization of data from a diverse set of domains including material science~\cite{GDNPBHH07, petruzza2019high}, cosmology~\cite{shivashankar2015felix}, and chemistry~\cite{gunther2014characterizing, bhatia2018topoms} have clearly demonstrated the utility of this topological structure. A sound theoretical framework for identification of features, a principled approach to measuring the size of features, controlled simplification, and support for noise removal are key reasons for the wide use of this topological structure.

Satisfying the interactivity requirement in feature-driven analysis and visualization has increasingly become a challenge due to the availability of higher precision and feature-rich data. Time-varying data  poses another challenge where each time step may have to be analyzed within a short time budget. Different stages of the analysis pipeline are optimized for runtime performance~\cite{peterka2011scalable} with the consequence that the computation of the MS complex is a computational bottleneck. Naturally, several methods proposed during the past decade for computing the MS complex employ parallel algorithms. These methods are all designed to execute on multicore CPU architectures, with the exception of a few methods where the embarrassingly parallel critical point computation step executes on the GPU.  In this paper, we present a fast parallel algorithm that computes the graph structure of the MS complex via a novel transformation to matrix and vector operations. It further leverages data parallel primitives resulting in an efficient end-to-end GPU implementation. We also present a CPU based parallel algorithm for simplifying the MS complex via iterative cancellations of critical point pairs.

\subsection{Related Work}
The development of effective workflows for the analysis of large scientific data based on the MS complex, coupled with increasing compute power of modern shared-memory and massively parallel architectures has generated a lot of interest in fast and memory efficient parallel algorithms for the computation of 3D MS complexes. Gyulassy et al.~\cite{gyulassy2008practical} introduced a memory efficient computation of 3D MS complexes, where they handled large datasets that do not fit in memory. Their method partitions the data into blocks, called parcels, that fit in memory. Next, it computes the MS complex for the individual parcels and uses a subsequent cancellation based merging of individual parcels to compute the MS complex of the union of the parcels. This framework was extended in the design of distributed memory parallel algorithms developed by Peterka et~al.~ \cite{peterka2011scalable} and Gyulassy et~al. ~\cite{gyulassy2012parallel}, where they additionally leverage high performance computing clusters to process the parcels in parallel. 
Subsequent improvements to parallelization were based on novel locally independent definitions for gradient pairs by Robins et~al.~\cite{robins2011theory} and Shivashankar and Natarajan~\cite{shivashankar3d} that allowed for embarrassingly parallel approaches for gradient assignment. Further improvements in computation time were achieved by efficient traversal algorithms for the extraction of ascending and descending manifolds of the extrema and saddles ~\cite{gunther2012efficient, shivashankar3d, delgado2014skeletonization}. 

Graph traversals for computing the ascending and descending manifolds of extrema, owing to their relatively simple structure, have been modeled as root finding operations in a tree and subsequently parallelized on the GPU~\cite{shivashankar3d}. However, the best known  algorithms for the more complex saddle-saddle traversals are still variants of a fast serial breadth first search traversal. More recently, Gyulassy et~al.~\cite{gyulassy2018shared, gyulassy2014conforming, gyulassy2012computing} and Bhatia et~al.~\cite{bhatia2018topoms} have presented methods that ensure accurate geometry while computing the 3D MS complex in parallel. The approaches described above, with the exception of Shivashankar and Natarajan~\cite{shivashankar3d}, implement CPU based shared memory parallelization strategies. Shivashankar and Natarajan~\cite{shivashankar3d} describe a hybrid approach and demonstrate the advantage of leveraging the many core architecture of the GPU. Embarrassingly parallel tasks such as gradient assignment and extrema traversals were executed on the GPU, resulting in a speedup over CPU based approaches. 

Some of the above-mentioned methods employ a discrete Morse theory based approach for defining and computing the MS complex with a focus on 2D and 3D scalar functions~\cite{gyulassy2006simplification,gunther2012efficient,delgado2014skeletonization,shivashankar2011parallel,shivashankar3d}. This approach results in combinatorial and numerically robust algorithms. The discrete Morse theory based approach is also amenable for extension to higher dimensions, as shown in recent work by Fugacci et al.~\cite{fugacci2019computing} who compute Morse complexes from simplicial complexes. The resulting discrete Morse complex finds applications to homology computation~\cite{robins2011theory,harker2014discrete} and analysis of shape and scalar fields~\cite{defloriani2015morse}.

\subsection{Contributions}
In this paper, we describe a fast GPU parallel algorithm for computing the MS complex. The algorithm employs the discrete Morse theory based approach, where the gradient flow is discretized to elements of the input grid. Key contributions include
\begin{itemize}
    \item An algorithm that utilizes fine grained parallelism for all steps of the MS complex computation and hence enables an efficient end-to-end GPU implementation. 
    \item Two novel ideas for transforming graph traversal into operations that are amenable to parallel computation: (a)~BFS tree traversal for determining saddle-saddle reachability is transformed into a sequence of vector operations, and (b)~saddle-saddle path computation is modeled as wave propagation and transformed into a sequence of matrix multiplication operations. These transformations help resolve a major computational bottleneck in previous parallel algorithms.
    \item Efficient parallel methods for populating the MS complex data structure.
    \item A data parallel algorithm for simplifying the MS complex. The algorithm utilizes a grid subdivision scheme to identify pairs of critical points that may be cancelled in parallel and results in improved runtimes and comparable quality of the simplified complex.
    \item A GPU implementation of the parallel algorithm, \gmsc, which is up to 8.6 times faster than pyms3d~\cite{shivashankar3d,mscsoftware2017,mscomplexsoftware} and 6 times faster than TTK~\cite{tierny2017topology}, which are existing methods. \gmsc uses highly optimized data parallel primitives such as prefix scan and stream compaction extensively, thereby leading to high scalability and efficiency.
    \item A detailed experimental analysis of the MS complex computation algorithm based on a study of the contributions of each step towards overall performance improvement, statistics that indicate the available parallelism for the different steps and how it is leveraged, and finally demonstrating how the different computational bottlenecks are removed via the transformations.
\end{itemize}

A previous paper presented a brief description of the GPU algorithm together with preliminary results~\cite{subhash2020GPU}. In this extended version, we additionally present methods for parallel population of the MS complex data structures, a parallel simplification algorithm, and improvements to the GPU implementation that results in better runtimes. We also perform extensive computational experiments on a large number of datasets and include a report of a detailed investigation of the different steps of the algorithm and their contribution towards the overall runtimes.

\section{Preliminaries}
\begin{figure}[htb]
   \centering
   \includegraphics[width=0.8\linewidth]{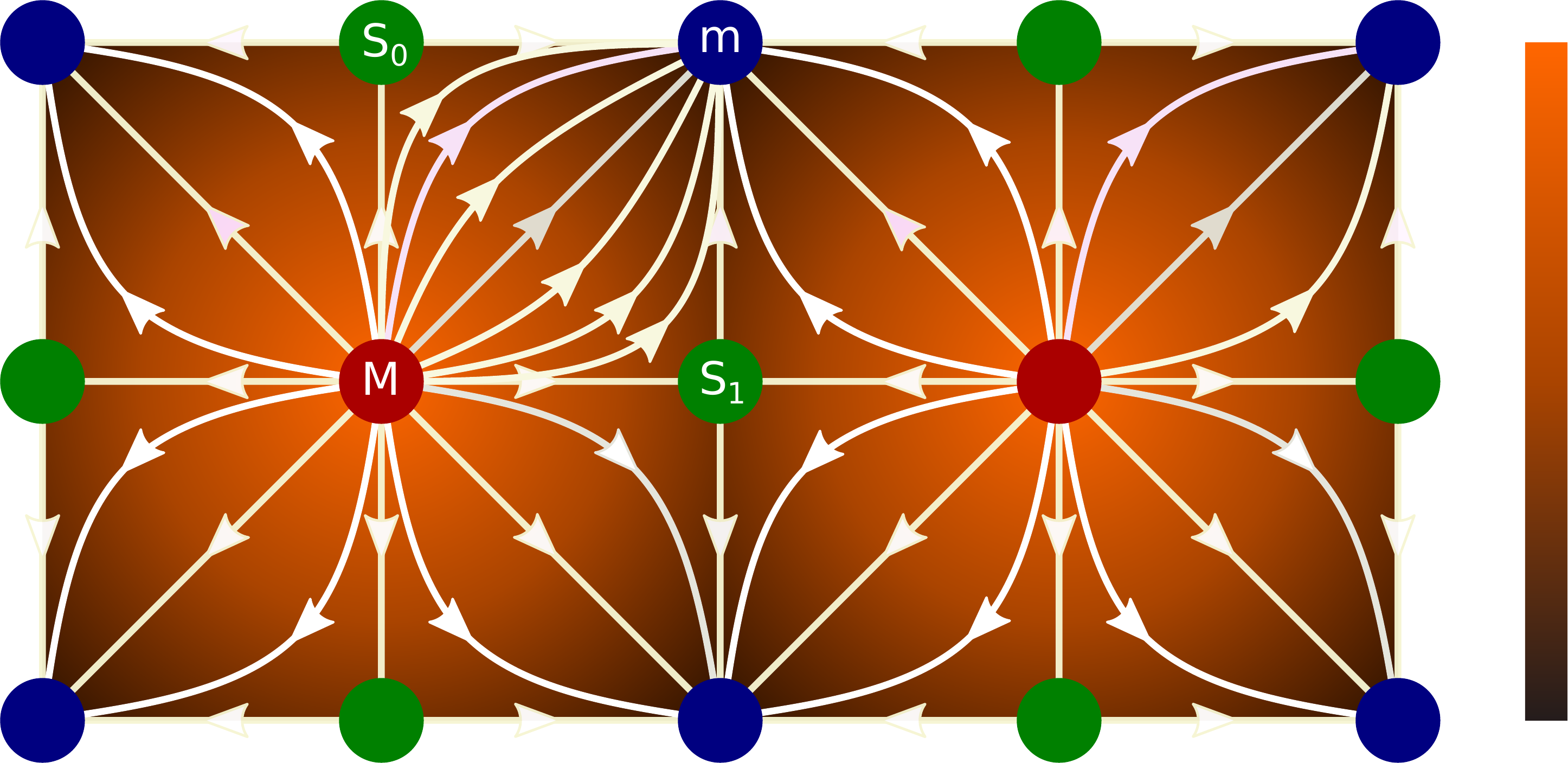}
     \caption{A 2D scalar function and its critical points (maxima - red, minima - blue, saddle - green) and reversed integral lines. A 2-cell $MS_0mS_1$ of the MS complex is a collection of all integral lines between $m$ and $M$. A 1-cell (say, $mS_1$ or $S_1M$) of the MS complex consists of the integral line between a minimum and a saddle or the integral line between a saddle and a maximum.}
     \label{fig:2D-mscomplex-illustration}
\end{figure}
\begin{figure*}[!t]
   \centering
   \includegraphics[width=1\linewidth]{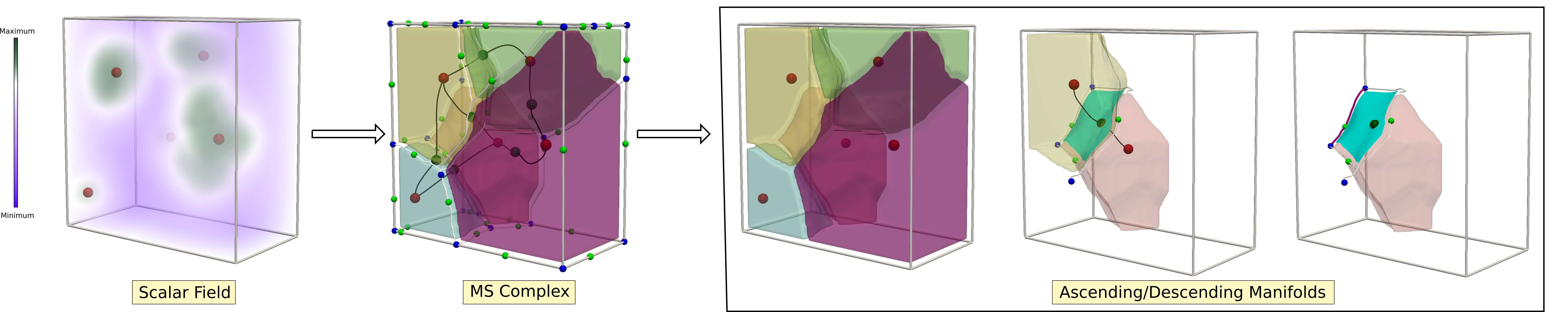}
    \caption{The 3D Morse-Smale complex is an abstract representation of the gradient flow of a 3D scalar function. (left to right) A scalar function; combinatorial structure of the corresponding MS complex consists of critical points and arcs connecting them; the descending 3-manifolds of maxima (red) that consist of integral lines that terminate at the corresponding maxima, descending 2-manifold of a 2-saddle (dark green) and its dual ascending 1-manifold, and the descending 1-manifold of a 1-saddle (light green).} 
    \label{fig:3D-mscomplex-illustration}
\end{figure*}
In this section, we briefly introduce the necessary  background on Morse functions and discrete Morse theory that is required to understand the algorithm for computing the 3D MS complex. 

\subsection{Morse-Smale Complex}
Given a smooth scalar function $f:\mathbb{R}^3\to\mathbb R$, the MS complex of $f$ is a partition of $\mathbb{R}^3$ based on the induced gradient flow of $f$.  
A point $p_c$ is called a \textit{critical point} of $f$ if the gradient of $f$ at $p_c$ vanishes, $\nabla f(p_c) = 0$. If the Hessian matrix of $f$ is non-singular at its critical points, the critical points can be classified based on their \textit{Morse index}, defined as the number of negative eigenvalues of the Hessian. Minima, 1-saddles, 2-saddles and maxima are critical points with index  equal to 0,1,2, and 3 respectively. An \textit{integral line} is a maximal curve in $\mathbb{R}^3$ whose tangent vector agrees with the gradient of $f$ at each point in the curve. The origin and destination of an integral line are critical points of $f$. The set of all integral lines originating at a critical point $p_c$ together with $p_c$ is called the \textit{ascending manifold} of $p_c$. Similarly, the set of all integral lines that share a common destination $p_c$ together with $p_c$ is called the \textit{descending manifold} of $p_c$.  

The \emph{Morse-Smale (MS) complex} is a partition of the domain of $f$ into cells formed by a collection of integral lines that share a common origin and destination. See \autoref{fig:2D-mscomplex-illustration} and \autoref{fig:3D-mscomplex-illustration} for examples in 2D and 3D. The cells of the MS complex may also be described as the simply connected cells formed by the intersection of the ascending and descending manifolds. The \emph{1-skeleton} of the MS complex consists of \emph{nodes} corresponding to the critical points of $f$ together with the \emph{arcs} that connect them. The 1-skeleton is often referred to as the combinatorial structure of the MS complex.
\begin{figure}[H]
   \centering
   \includegraphics[width=1\linewidth]{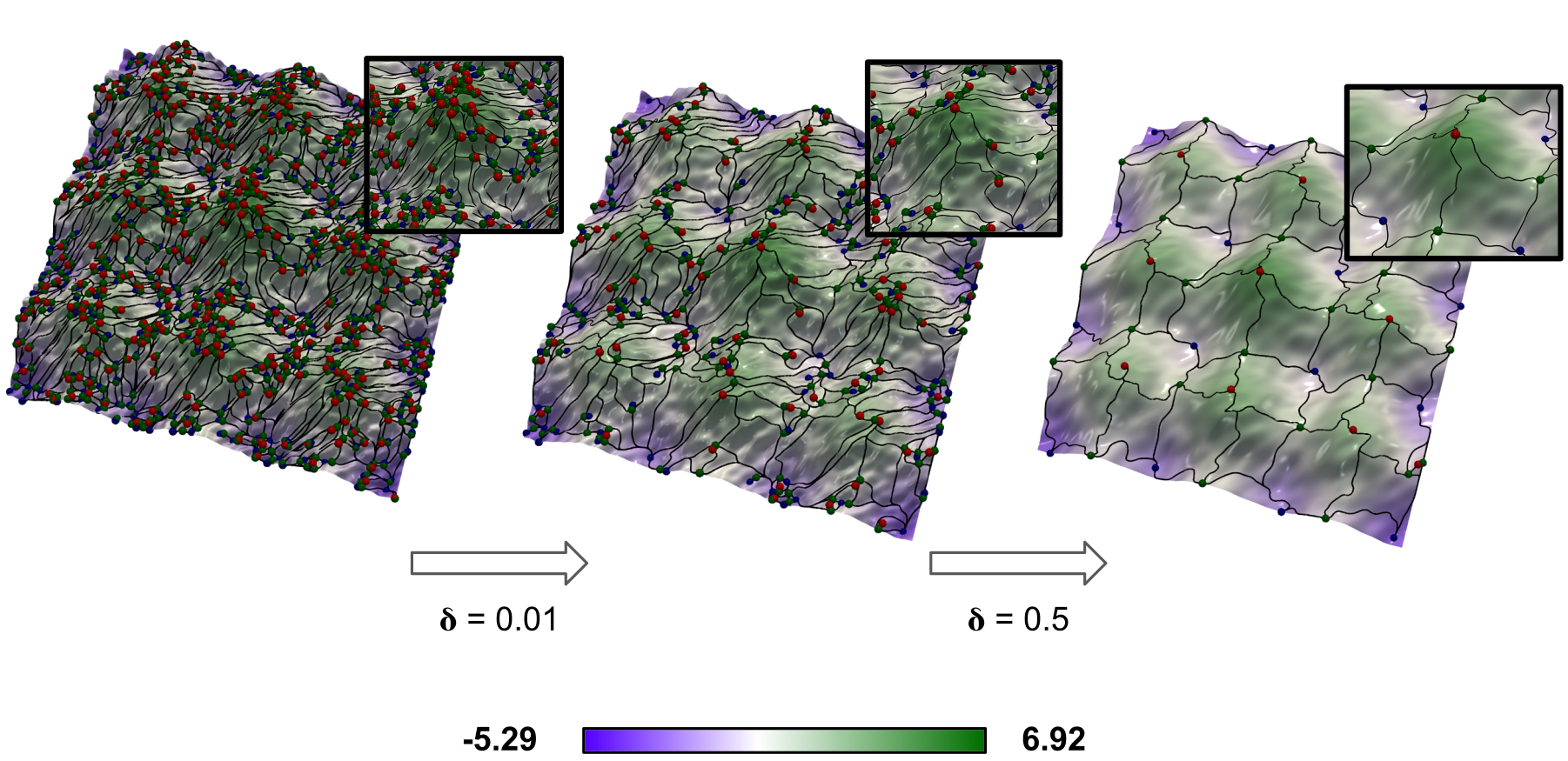}
    \caption{Persistence ($\delta$) driven simplification of a 2D MS complex. At a threshold of 0.01, all pairs of critical points (saddle-maxima or saddle-minima in 2D) whose persistence is below 0.01 are cancelled. Increasing the threshold further to 0.5 removes all small sized topological features revealing the significant mountains, represented by the corresponding peaks. Note that when a maximum is cancelled, arcs that terminate at that critical point are extended to the surviving maximum.} 
    \label{fig:simplification}
\end{figure}

The scalar function may be simplified via repeated \emph{cancellation} of critical point pairs that are connected by a single arc in the MS complex, see \autoref{fig:simplification}. A cancellation operation removes the critical point pair, the arc between them, and reconnects the neighbors in the 1-skeleton. A cancellation corresponds to a local smoothing of the scalar function~\cite{EHZ03}. The difference in function value between the critical point pair is a good measure of the effect of cancellation on the function. In 2D, the least persistent~\cite{edelsbrunner2000persistence} critical point pair is connected by an arc in the MS complex~\cite{EHZ03}. MS complex simplification via repeated critical point pair cancellation is often referred to as persistence-driven simplification due to this relationship between the measure of an arc and topological persistence. Cancellations ordered based on the difference in function value between the critical point pair result in a natural sequence of simpler MS complexes.

\subsection{Discrete Morse Theory}
\begin{figure}[htb]
   \centering
   \includegraphics[width=1\linewidth]{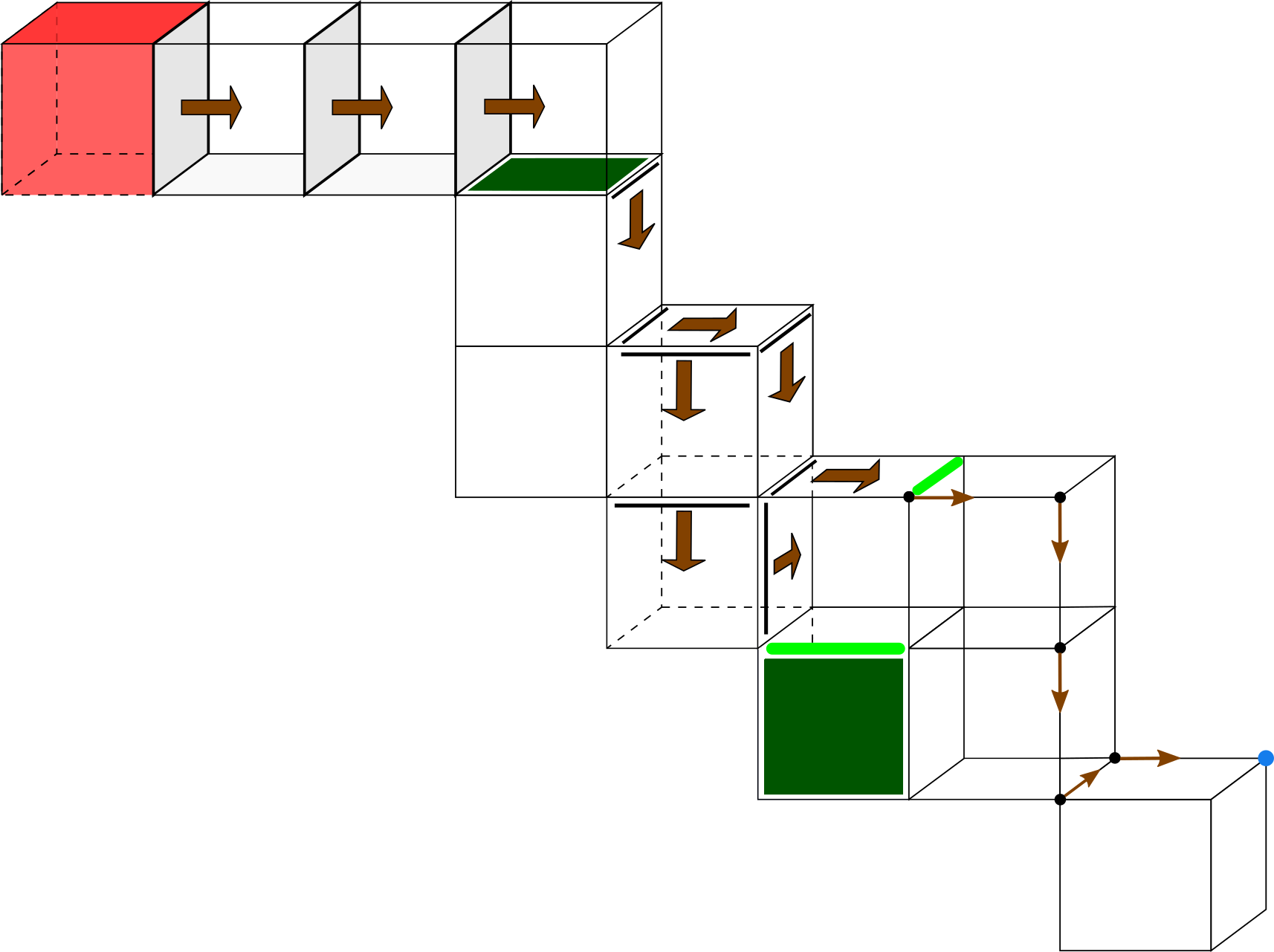}
    \caption{Discrete gradient, critical cells, and gradient paths for a function defined on a grid. The figure shows a subset of the grid. Gradient pairs (brown arrows) between two cells constitute gradient paths. A cell that is not paired is critical. There are four types of critical cells: maximum (red), 2-saddle (dark green), 1-saddle (light green), and minimum (blue).} 
    \label{fig:discrete-grid}
\end{figure}
Discrete Morse theory, introduced by Forman \cite{forman2002user}, is a combinatorial analogue of Morse theory that is based on the analysis of a discrete gradient vector field defined on the elements of a cell complex. 
Adopting an approach based on discrete Morse theory for computing the MS complex results in robust and computationally efficient methods with the added advantage of guarantees of topological consistency~\cite{gyulassy2008practical,robins2011theory, shivashankar2011parallel,gunther2012efficient,shivashankar3d,shivashankar2014}. 
We focus on scalar functions defined on a 3D grid, a cell complex with cells of dimensions 0,1,2, or 3. If an $i$-cell $\alpha$ is incident on an $(i+1)$-cell $\beta$ then $\alpha$ is called a \emph{facet} of $\beta$ and $\beta$ is called a \emph{cofacet} of $\alpha$.

A \textit{gradient pair} is a pairing of two cells $\langle \alpha^{(i)},\beta^{(i+1)} \rangle$, where $\alpha$ is a facet of $\beta$, see \autoref{fig:discrete-grid}. The gradient pair is a discrete vector directed from the lower dimensional cell to the higher dimensional cell. It corresponds to the negative gradient at a point for smooth functions. A \textit{discrete vector field} defined on the grid is a set of gradient pairs where each cell of the grid appears in at most one pair. A \textit{critical cell} with respect to a discrete vector field is one that does not appear in any gradient pair. The \emph{index} of the critical point is equal to its dimension. A \textit{V-path} in a given discrete vector field is a sequence of cells $\alpha_0^{(i)},\beta_0^{(i+1)},\alpha_1^{(i)},\beta_1^{(i+1)}....,\alpha_r^{(i)},\beta_r^{(i+1)},\alpha_{r+1}^{(i)}$ such that $\alpha_k^{(i)}$  and $\alpha_{k+1}^{(i)}$ are facets of $\beta_k^{(i+1)}$ and $\langle \alpha_k^{(i)},\beta_k^{(i+1)} \rangle$ is a gradient pair for all $k=0..r$. A V-path is called a \textit{gradient path} if it has no cycles and a \textit{discrete gradient field} is a discrete vector field which contains no non-trivial closed V-paths. A gradient path is maximal if there exists no longer gradient path that contains it. Maximal gradient paths for a function defined on a grid correspond to the notion of integral lines in the smooth setting, see \autoref{fig:msc-max-min-dag}. We can thus similarly define ascending and descending manifolds for discrete scalar functions defined on a grid. 

\subsection{Parallel Primitives}
Data parallel primitives serve as effective tools and building blocks in the design of parallel algorithms. We leverage them in all steps of the MS complex computational pipeline. Specifically, we use two primitives, \emph{prefix scan} and \emph{stream compaction}.
\begin{figure}[htb]
   \centering
   \includegraphics[width=1.0\linewidth]{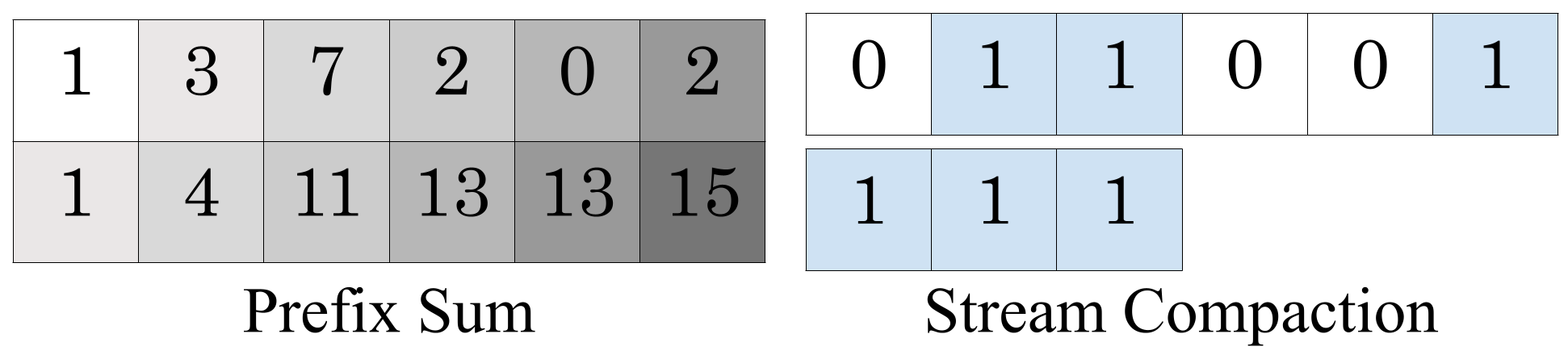}
     \caption{(left)~A prefix sum operator applied on the top array produces the array on the bottom. (right)~The boolean array on top is stream compacted to generate an array of true boolean values.}
     \label{fig:dataparallelprimitives}
\end{figure}

Given an array of elements and a binary reduction operator, a prefix scan is defined as an operation where each array element is recomputed to be the reduction of all earlier elements~\cite{harris2007parallel}. If the reduction operator is addition, the operator is called \textit{prefix sum}. Given an array of elements, stream compaction produces a reduced array consisting of elements that satisfy a given criterion. \autoref{fig:dataparallelprimitives} illustrates the two operators. We use Thrust~\cite{thrustWeb}, a popular library packaged with CUDA~\cite{cudaWeb}, for the prefix scan and stream compaction routines.

\begin{figure*}[htb]
   \centering
   \includegraphics[width=1\linewidth]{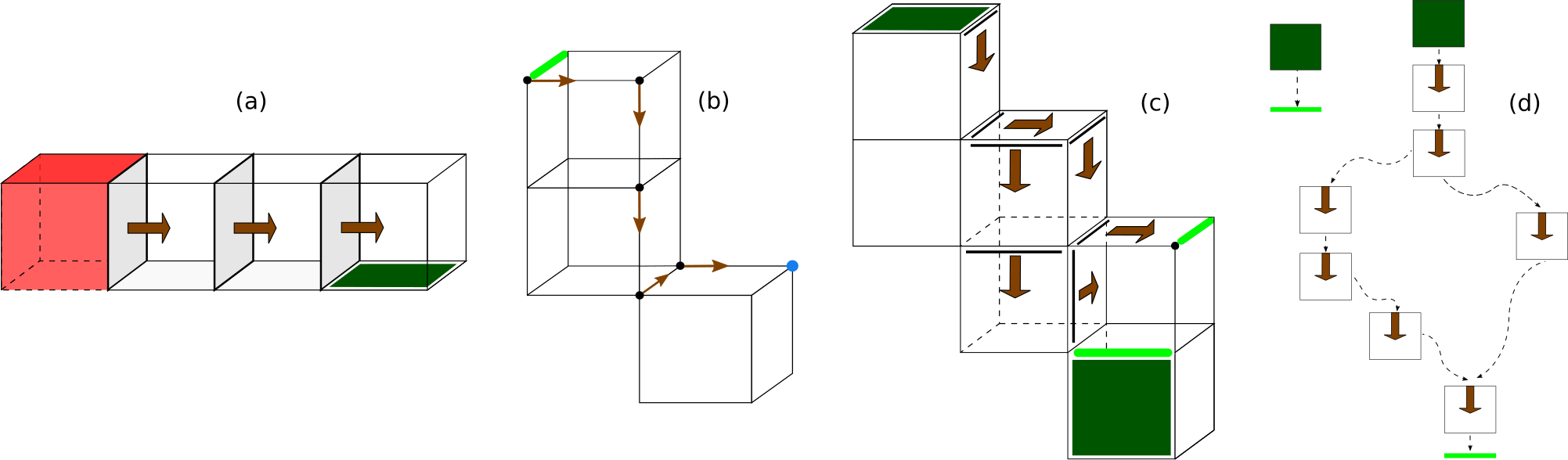}
   \caption{Maximal gradient paths between (a)~maximum (red) and a 2-saddle (dark green), (b)~1-saddle (light green) and a minimum (blue), and (c)~2-saddle and a 1-saddle.  (d)~A directed acyclic graph~(DAG) representing the collection of gradient paths between 2-saddles and 1-saddles. Dashed edges of the DAG represent incidence relationship between a 2-saddle or the 2-cell of a (1,2) gradient pair and the 1-cell of a (1,2) gradient pair or a 1-saddle. There exist two possible gradient paths between the 2-saddle and 1-saddle due to the presence of a split and merge. A sequence of such split-merge configurations results in an exponential increase in the number of gradient paths. The DAG contains an additional component consisting of only two nodes in the special case when a 2-saddle is adjacent to a 1-saddle (bottom cube in (c)).} 
    \label{fig:msc-max-min-dag}
\end{figure*}

\section{Algorithm}
Discrete Morse theory based algorithms for computing the MS complex on a regular 3D grid consist of two main steps~\cite{gyulassy2008practical,robins2011theory,gunther2012efficient,shivashankar3d}. Both steps are amenable to parallel computation. The first step computes a well defined discrete gradient field and uses it to locate all critical cells. Each 3-cell may be processed in parallel together with its lower dimensional cells. The second step computes ascending and descending manifolds as a collection of gradient paths that originate and terminate at critical cells.  A combinatorial connection between two critical cells is established if there exists a gradient path between them. This step may be parallelized by assigning the source of each ascending and descending manifold to a different thread. The collection of critical cells together with the connections is stored as the combinatorial structure of the MS complex. 



The second step computes two types of gradient paths: saddle-extrema and saddle-saddle paths. Different traversal strategies have been proposed to compute these paths. The parallel algorithm with the best known performance for shared memory processors employs strategies that leverage the characteristics of the gradient path structure in the two cases~\cite{shivashankar3d}. They compute the saddle-extremum paths using root finding operations on a tree. The saddle-saddle path computation is reduced to a multi-source multi-destination path counting problem in a directed acyclic graph. This path counting is affected by an exponential increase in the number of  paths between saddles due to the presence of multiple splits and merges in gradient paths. All paths originating from a source are computed by the same thread, thereby serializing the computation. Thus, saddle-saddle path computation is the primary performance bottleneck in current parallel algorithms for MS complex computation, and we address it in our algorithm.

We broadly follow the discrete Morse theory based approach employed by Shivashankar and Natarajan~\cite{shivashankar3d} for computing the MS complex. We follow their approach in the first step that identifies critical cells, and for the saddle-extrema path computation in the second step. However, we introduce new algorithms for computing the gradient paths between saddles. Specifically, we introduce novel transformations that reduce the saddle-saddle gradient path traversal into highly parallelizable vector and matrix operations. These transformations utilize ideas for computing paths in graphs via matrix operations~\cite{seidel1995all,zwick2001exact,zwick2002all}.
 
In this section, we restrict the description to the newly developed algorithms and refer the reader to earlier work~\cite{shivashankar3d} for details on critical cell identification and saddle-extrema arc computation. Our implementation of the critical cell and saddle-extrema arc computation additionally incorporates improved methods for storage using data parallel primitives, as described in \autoref{sec:implementation-optimization}.

\subsection{1-Saddle -- 2-Saddle connections}
Gradient paths between 1-saddles and 2-saddles can both merge and split as shown in \autoref{fig:msc-max-min-dag}~(c). A sequence of merges and splits result in an exponential growth in the number of gradient paths. Computing the saddle connections is the major computational bottleneck in current methods for computing the MS complex. The saddle-saddle arc computation is analogous to counting paths between multiple sources and destinations in a directed acyclic graph (DAG) that represents all 1-2 gradient pairs (\autoref{fig:msc-max-min-dag}~(d)). A trivial task based parallelization across all source saddles is affected by the exponential number of paths that need to be explored by each thread and hence an increase in storage requirement. Existing approaches tackle this problem by first marking reachable pairs of saddles using a serial BFS traversal. Next, they traverse paths between reachable pairs and count all paths using efficient serial traversal techniques.

In the following subsections, we introduce transformations that enable efficient parallel computation of both steps, reachability computation and gradient path counting. The reachable pairs in the DAG are identified using a fine-grained parallel BFS algorithm~\cite{merrill2012scalable} that is further optimized for graphs with bounded degree. In order to compute all gradient paths between reachable pairs, we first construct a minor of the DAG by contracting all simple paths to edges. We represent the minor using adjacency matrices and compute all possible paths between the reachable pairs via a sequence of matrix multiplication operations. These transformations enable scalable and fine grained parallel implementations of both steps without the use of synchronization or locks.

\subsection{Saddle-Saddle reachability}
We mark reachable saddle-saddle pairs using parallel BFS traversals of the DAG starting at all 1-saddles. A BFS traversal iteratively computes and stores a frontier of nodes that are reachable from the source node. The frontier is initialized to the set of all source nodes, namely the 1-saddles. After the $i^{th}$ iteration, the frontier consists of nodes reachable via a gradient path of length $i$. We observe from Figure~\ref{fig:msc-max-min-dag}~(c) that while traversing a gradient path, the next 1-2 gradient pair should necessarily contain one of four possible 1-cell facets of the 2-cell from the current 1-2 gradient pair. So, the outdegree of a node in the DAG is at most 4, which in turn imposes a bound on the size of the frontier in the next iteration. This bound allows us to allocate sufficient space to store the next frontier. All nodes in the current frontier are processed in parallel to update the frontier. Subsequently, a stream compaction is performed on the frontier to collect all nodes belonging to the next frontier. The traversal stops when it reaches a 2-saddle and the algorithm terminates when the frontier is empty. 

\begin{figure}[htb]
   \centering
   \includegraphics[scale=0.45]{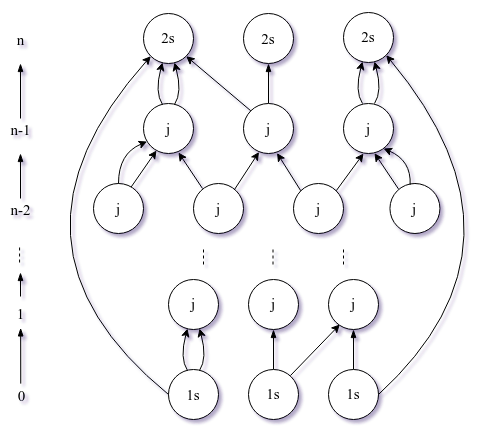}
     \caption{\label{fig:condensed-graph}
           Gradient path counting in the DAG minor. We count paths between 1-saddles and 2-saddles in the DAG by iteratively identifying paths of increasing length in the DAG minor. The $i^{th}$ iteration identifies all paths of length $i$ in the DAG minor, discovering the paths from nodes in level-0 to nodes in level-$i$.} A node in level-$i$ may be discovered again in a later iteration. In this case, the algorithm records both paths. 

\end{figure}

\subsection{Gradient path counting}
We now discuss an efficient parallel algorithm for traversing and counting all gradient paths between the reachable pairs identified earlier. So, we restrict our attention to the subgraph of the 1-skeleton of the MS complex consisting of gradient paths between 1-saddles and 2-saddles. Edges belonging to the corresponding subgraph of the DAG are marked during the previous BFS traversal step. Let $G$ denote this subgraph. Direct traversal and counting the number of paths between a pair of reachable saddles is again  inherently serial. Instead, we first construct a minor $G'$ of $G$ by contracting all simple sub-paths in $G$.  This transformation  results in a graph $G'$ whose nodes are either 1-saddles, 2-saddles, or junction nodes. A \emph{junction node} is a node whose outdegree is greater than 1. A path in the graph $G$ is \emph{simple} if none of its interior nodes is a junction node. The end points of a maximal simple path of $G$ is either a junction node or a saddle. We construct $G'$ by traversing all maximal simple paths of $G$ in parallel. Paths in $G'$ are counted using a sequence of matrix multiplication operations, which are amenable to fine grained parallelism.

\myparagraph{DAG minor construction.}
 We compute a minor $G'$ of $G$ by contracting all simple paths between 1-saddles, 2-saddles and junction nodes into edges between their source and destination nodes. We process all 1-saddles and junction nodes in parallel to trace all paths that originate at that node. The paths terminate when they reach a junction node or a 2-saddle. So, all paths are guaranteed to reach their destination without splits. Note that some of the paths may merge. However, such paths originate from different source nodes and are processed by different threads. 

All path traversals share a common array to record the most recent visited node in each path. In each iteration, all paths advance by a single node. The paths terminate if they reach a 2-saddle or a junction node. We perform a stream compaction on the array at the end of the iteration to remove terminated paths and to ensure that all threads are doing useful work. The algorithm terminates when all paths have been marked as terminated. 
When a path terminates, the thread inserts an edge into an adjacency list representation of $G'$. Two different paths between a pair of junction nodes or between a saddle and junction node are recorded as two different copies of the edge. The adjacency list is converted into a sparse matrix in the next step, at which time copies of edges in $G'$ are stored as edge multiplicity.

%
%

\myparagraph{Path counting via matrix operations.}
Nodes of the graph minor $G'$ constructed via path contraction may be placed in levels as shown in \autoref{fig:condensed-graph}.  
We convert the adjacency list representation of $G'$ into sparse matrices. This conversion is crucial because it enables the counting of the number of paths between all 1-saddle -- 2-saddle pairs using iterative matrix multiplication. Edges in $G'$ may be classified into four different types: $1s-j$, $j-j$, $j-2s$ and $1s-2s$.  Let $A$\textsubscript{$1s-j$}, $B$\textsubscript{$j-j$}, $B$\textsuperscript{*}\textsubscript{$j-2s$}, and $D$\textsubscript{$1s-2s$} represent the adjacency matrices that store the respective edges with multiplicity. For example, rows of the adjacency matrix $A$\textsubscript{$1s-j$} correspond to 1-saddles, columns correspond to the junction nodes, and the matrix elements denote the multiplicity of the $1s-j$ edge in $G'$.
%

Consider nodes belonging to consecutive levels in \autoref{fig:condensed-graph}, in particular between level-0 and level-1. We first multiply $A$\textsubscript{$1s-j$} $\times$ $B$\textsubscript{$j-j$} and store the result in $A$\textsubscript{$1s-j$}, which establishes connections between level-0 and level-2 nodes with multiplicity. Each successive multiplication of $A$\textsubscript{$1s-j$} with $B$\textsubscript{$j-j$} establishes connections between level-0 and the newly discovered level.  A simple sequence of these matrix operations may not result in the desired set of path traversals. In particular, if a node is revisited at multiple levels, it appears at different depths from the source and each unique visit should be recorded. We ensure that such nodes are also handled correctly by updating the sequence of matrix operations so that it maintains a storage matrix $A\textsuperscript{*}$\textsubscript{$1s-j$} that collects all newly discovered paths after each iteration. We describe the procedure in detail below.

Initialize four matrices with their respective edges as follows: $A$\textsubscript{$1s-j$}, $B$\textsubscript{$j-j$}, $B\textsuperscript{*}$\textsubscript{$j-2s$}, $D$\textsubscript{$1s-2s$}. We maintain $A\textsuperscript{*}$\textsubscript{$1s-j$}, which iteratively records newly discovered paths. Each iteration multiplies $A$\textsubscript{$1s-j$} with $B$\textsubscript{$j-j$} to yield a matrix $C$\textsubscript{$1s-j$}, after which we add matrix $A$\textsubscript{$1s-j$} to $A\textsuperscript{*}$\textsubscript{$1s-j$}. The newly discovered paths in $C$\textsubscript{$1s-j$} become the input frontier $A$\textsubscript{$1s-j$} for the next iteration’s multiplication. 

After the final iteration, $A\textsuperscript{*}$\textsubscript{$1s-j$} contains all possible paths from 1-saddles to junction nodes. The purpose of $A\textsuperscript{*}$\textsubscript{$1s-j$} is two-fold: first, it records shorter paths that terminate during intermediate iterations. 
Next, nodes that are discovered multiple times add their incrementally discovered paths to it after each multiplication. This matrix is then multiplied by $B\textsuperscript{*}$\textsubscript{$j-2s$} to obtain $D\textsuperscript{*}$\textsubscript{$1s-2s$}, establishing connections between 1-saddles and 2-saddles. Edges between 1-saddles and 2-saddles in $G'$ are stored in  $D$\textsubscript{$1s-2s$}. In the final step,  $D$\textsubscript{$1s-2s$} is added to $D\textsuperscript{*}$\textsubscript{$1s-2s$} to record the set of paths with multiplicity between all 1-saddles and 2-saddles. Matrices are stored in their sparse forms, thereby considerably reducing the memory footprint, see Section 3.5. We use the CUDA sparse matrix library cu\textsc{SPARSE}~\cite{cuSparseWeb} for matrix multiplication operations.

\subsection{Saddle-Extrema arcs}
\label{sec:saddle-extrema-arcs}
The difference in the structure of gradient paths that originate / terminate at extrema versus those originating / terminating at saddles necessitates different approaches for their traversals.  Gradient paths that originate or terminate at extrema can either split or merge but not both. So, traversing saddle-extrema gradient paths is analogous to a root finding operation in a tree.
We employ a parallel tree traversal algorithm that computes the saddle-extrema arcs while also computing the descending manifolds of maxima and ascending manifolds of minima~\cite{shivashankar3d}. The algorithm traverses gradient paths from all cells in parallel via pointer jumping~\cite{jaja1992parallelalgorithms}, where each cell's destination is updated in every iteration. This algorithm performs well in practice, so we employ the same in our computation. However, we incorporate code optimizations as described in the following section to further improve the runtimes of the routines that populate the data structures.
 
\subsection{Implementation and optimizations}
\label{sec:implementation-optimization}
\gmsc implements the algorithm in C++ and uses the CUDA framework for parallel computation on the GPU in contrast to the OpenCL based implementation in pyms3d~\cite{shivashankar3d,mscsoftware2017,mscomplexsoftware}. In this section, we describe novel implementation level optimizations in \gmsc that aim to improve performance in terms of running time, data structure population, and memory usage. 

We store the ascending and descending 1-manifolds as two directed graphs whose nodes are the critical points of the scalar function. Each graph is stored as an STL vector of maps, to efficiently retrieve paths in either direction on demand. Each element of the STL vector represents a critical point and the corresponding map stores the adjacency list of the critical point. We populate the graph using a pre-sized parallelizable vector, coupled with GPU parallel prefix sum and stream compaction routines that enable efficient data handling.

\myparagraph{Gradient pairs and critical points.}
Gradient pairs are computed in the first step of the algorithm. The scalar field is sampled at all vertices of the 3D grid and the gradient pair information is stored at each cell of the grid as a bit vector. The gradient pairs and scalar fields are stored in specialized memory on the GPU and are accessed using \emph{texture and surface objects} provided by CUDA. This surface and texture memory facilitates high performance and bandwidth through spatial locality in the reads~\cite{cudaWeb}. Further, they also serve as a cache. \gmsc leverages this option by maximizing coalesced memory access as opposed to a traditional global memory usage. 

The critical points are identified and marked in the second step. In order to count them efficiently, we employ parallel reduction techniques in conjunction with shared memory synchronization to obtain block-wise sums. This is followed by a prefix sum from the Thrust library~\cite{thrustWeb} which gives us the total number of critical points. We also use Thrust's stream compaction for optimized population of critical point information in memory, by marking critical cells and discarding the rest using compaction.

\myparagraph{Saddle-Saddle reachability.}
In contrast to a serial BFS based reachability computation~\cite{shivashankar3d}, \gmsc marks all reachable paths between 1-saddles and 2-saddles in parallel. It collects all 1-saddles by marking them in parallel using boolean values and establishes their ordering using a prefix sum. It uses Thrust device vectors for dynamic memory allocation and to store the current and discovered frontiers in each BFS iteration. The size of the discovered frontier can be at most four times the size of the current frontier. The first frontier will be the set of 1-saddles and each thread is assigned to explore its designated saddle's cofacet pairs. All discovered 1-2 pairs constitute paths leading towards 2-saddles. We mark these pairs as visited and store them in the discovered frontier. Invalid pairs are also marked. A stream compaction collects all valid discovered pairs, copies them to the next iteration's current frontier and discards the rest. When all paths reach 2-saddles, the compaction returns an empty set of discovered pairs and the traversal algorithm terminates.

\myparagraph{Gradient path counting.} 
The degree of parallelization obtained in the OpenMP based implementation of gradient path counting in pyms3d is limited by the number of available CPU cores and the number of source saddles. Our algorithm overcomes this limitation by transforming the computation into GPU parallelizable matrix operations. The dynamic resizing capabilities offered by Thrust device vectors prove to be an integral part of all subsequent steps. 
We compute the junction nodes in parallel by counting the number of cofacet pairs of each reachable 1-cell, except for 1-saddles, and marking it as a junction node if the outdegree is greater than 1. We further construct an ordered list of these junction nodes using prefix sum and stream compaction. 

The DAG minor is constructed by traversing, in parallel, the simple paths between saddles and junction nodes. The source-destination pairs may be stored in the adjacency matrices $A$\textsubscript{$1s-j$}, $B$\textsubscript{$j-j$}, $B$\textsuperscript{*}\textsubscript{$j-2s$}, and $D$\textsubscript{$1s-2s$}. At most four paths originate at a source node. So, we store the DAG minor using adjacency lists. Two arrays store the current frontier of the different paths and its stream compacted counterpart. The latter is processed in the subsequent iteration to obtain the next frontier. The path traversals are performed in two passes, with 1-saddles and junction nodes as sources, respectively. This partition reduces the memory footprint within each iteration since we store and populate only two adjacency lists at a time instead of four. All arrays are allocated as Thrust device vectors.
We ensure that the traversals, both in the parallel BFS and path counting routines, consistently move in the direction opposite to the discrete gradient in order to prevent cycles. Some paths do not terminate at a junction node or a 2-saddle. We discard such paths.

We convert the adjacency lists to the Compressed Sparse Row (CSR)~\cite{cuSparseWeb} representation of the adjacency matrix to enable efficient matrix operations. The CSR and CSC (Compressed Sparse Column) formats are popular and efficient representations of sparse matrices. Each iteration of the path counting algorithm requires a matrix multiplication followed by a matrix addition. Addition in the sparse form requires knowledge of the non-zero pattern in the matrix, which is stored in a temporary array before copying it to $A$\textsuperscript{*}\textsubscript{$1s-j$}. The iterations terminate when the multiplication returns a zero matrix, indicating that all $1s-j$ paths are stored in $A$\textsuperscript{*}\textsubscript{$1s-j$}. In the final step, the product of $A$\textsuperscript{*}\textsubscript{$1s-j$} and $B$\textsuperscript{*}\textsubscript{$j-2s$} is added to $D$\textsubscript{$1s-2s$} using a temporary matrix. The matrix $D$\textsubscript{$1s-2s$} is stored in both CSR and CSC formats, and is finally transferred to the main memory to enable efficient parallel storage of the combinatorial structure of the MS complex. Since the ascending and descending manifolds are stored as two directed graphs, the CSR and CSC formats can be directly used to populate their respective directions efficiently. We assign a thread to insert all paths originating at a given critical point in a given direction, thus achieving efficient parallel storage. 
We present experimental evidence of the sparsity of the matrices in \autoref{sec:experiments} that justifies the sparse representation. The matrix operations are implemented using the cuSPARSE library. Further, we reduce memory footprint at each step by effectively using Thrust’s stream compaction and dynamically allocated device vectors.

\myparagraph{Saddle-Extrema arcs.}
We implement the parallel tree traversal algorithm by~\cite{shivashankar3d} in CUDA. The destination of paths originating at each cell in the domain are updated iteratively using two buffers, implemented as 3D surface objects, the benefits of which were discussed earlier. Two buffers are required to avoid race conditions during parallel source and destination updates. We swap their values after each iteration. We also maintain a global boolean flag on the GPU accessed by all threads, which acts as an indicator for whether a path terminated by reaching its end point extremum. We check this flag after each iteration. If it is set, we verify if all paths have reached their destination extrema and terminate the algorithm accordingly. We choose Thrust device vectors in conjunction with prefix sum, stream compaction, and cuSPARSE conversion routines to efficiently store the final paths between all saddles and extrema in CSR and CSC formats. An additional array stores paths between all non-critical domain cells and their destination extrema. The traversals are performed in two successive passes, the first collecting 1-saddle -- minima and the second collecting 2-saddle -- maxima paths.

\myparagraph{Data structure.}
In order to facilitate efficient querying of the ascending and descending manifolds of the MS complex, we employ an STL vector to store critical points and individual maps for storing the ascending and descending arcs incident on each critical point. All arcs of the MS complex are stored twice, in a bidirectional manner, within this data structure. The data structure is populated in a serial step and is hence a computational bottleneck. We reduce the time required to populate this data structure by parallelizing the population across the STL vector of critical points. Each CPU thread populates the arcs incident on one critical point in a given direction. When the saddle-extrema and saddle-saddle paths are computed, the MS complex arcs are stored in both CSR and CSC formats, followed by a GPU--CPU data transfer operation. The tree traversal algorithm for computing the saddle-extrema arcs does not report the results in the desired CSR and CSC formats. So, we convert them explicitly.

\subsection{Runtime analysis}
Let $n$ denote the number of vertices in the input cube grid. Gradient pair and critical point computation takes $O(1)$ time per cell in the grid, and saddle-extrema arc computation requires $O(\log n)$ iterations following the algorithm from pyms3d~\cite{shivashankar3d}. The parallel BFS algorithm executes in $O(n)$ iterations since the depth of the tree can be $O(n)$. Parallel stream compaction at the end of each iteration requires $O(log(n))$  time~\cite{harris2007parallel} using the work efficient implementation in CUDA. So, parallel BFS for saddle-saddle reachability runs in $O(n\log n)$ time. The number of iterations in the DAG minor construction is $O(n)$. Accounting for stream compaction gives us a runtime complexity of $O(n\log n)$. All matrix multiplication and matrix addition operations in each iteration of the path counting algorithm are performed in parallel. Matrix multiplication takes $O(\log n)$ time and matrix addition can be performed in $O(1)$ time. The path counting algorithm executes in $O(n)$ iterations, corresponding to the length of the longest path between two nodes of the DAG minor. So, the total time taken for the MS complex computation is $O(n\log n)$.

\section{Parallel Simplification}
Parallel simplification is a relatively unexplored problem due to the intrinsic serial nature of the operation. Existing literature suggests that the ordering of critical point pair cancellations determines the runtime performance and final quality of the simplified scalar field~\cite{GDNPBHH07,gyulassy2008practical}. We propose a parallel algorithm that leverages a grid subdivision scheme and naturally imposes a specific sequence of cancellations. Our method achieves superior runtimes while maintaining comparable quality of the simplified MS complex with respect to the serial counterpart. 

\subsection{Grid subdivision}
We subdivide the underlying domain grid into regular subgrids and perform persistence ordered cancellations within each subgrid in parallel. The arcs are tested for independence prior to cancellation.
We determine if a given critical point pair in the 1-skeleton of the MS graph may be canceled by computing its arc multiplicity and persistence. Critical point pairs connected by multiple distinct arcs are not viable candidates for cancellation because they lead to a \emph{strangulation}~\cite{gyulassy2006simplification}. 
A valid cancellation of a critical point pair $pq$ ($index(p)$ = $index(q)+1$) modifies the 1-skeleton as follows: the descending arcs of $p$ are rerouted towards the end points of the ascending arcs of $q$, the nodes $p$ and $q$ and arcs incident on them are deleted. Clearly, the cancellation affects the neighborhood of the arc $pq$.  

\begin{figure}[htb]
   \centering
   \includegraphics[width=1.0\linewidth]{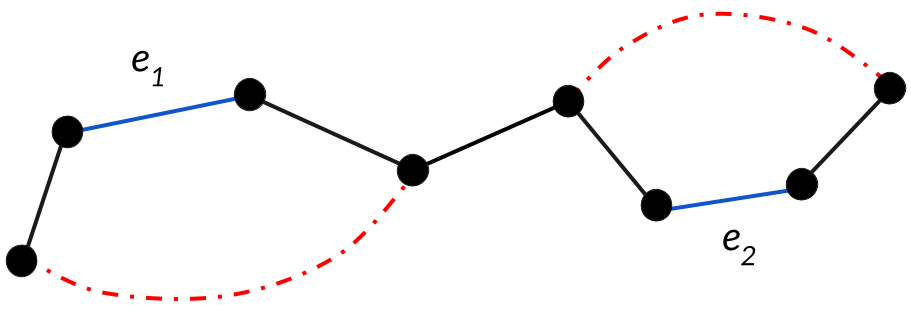}
     \caption{3-link neighborhood. Arcs $e_1$ and $e_2$ may be canceled concurrently because they do not lie within each other's 3-link neighborhood. Canceling $e_1$ and $e_2$ introduces the red reconnection arcs. No node is concurrently impacted by the two cancellations. Hence, all operations that update the 1-skeleton are thread safe.}
     \label{fig:3-link-nbd}
\end{figure}

Performing conflict-free cancellations in parallel (\ie without concurrent updates to a given node from two different cancellations) necessitates a minimum distance of three arcs between two critical point pairs (a 3-link neighborhood) as shown in \autoref{fig:3-link-nbd}. In order to cancel arcs that belong to two different subgrids in parallel, we grow a constraint region around the common boundary between the subgrids. This constraint region consists of arcs that are incident on or cross the common boundary between the subgrids and arcs that are adjacent to these boundary arcs. The constraint region is computed in parallel over all boundary arcs. 

The set of all arcs within each subgrid that do not belong to the constraint regions can be canceled serially. We leverage CPU-based parallelism (OpenMP) across all subgrids. The available parallelism, which is dictated by the number of subgrids, is insufficient to fully exploit GPU parallelism while offsetting the CPU-GPU data transfer costs. 

The simplification algorithm proceeds as follows. The domain grid is subdivided into subgrids via iterative slicing along the X, Y, and Z directions. Boundary arcs and constraint regions are marked in parallel. Next, we proceed with the cancellation in parallel across each subgrid. Each thread serially cancels all valid arcs contained within its designated subgrid boundaries and updates the MS complex concurrently. In order to ensure that cancellable arcs lying within the constraint regions are also scheduled for cancellation, we repeat the process by shifting the shared boundaries of subgrids using fixed offsets. We apply a final round of serial cancellation to handle low persistence arcs that may have escaped cancellation.

Let $n$ denote the number of vertices in the input cube grid and $m$ denote the maximum degree of a node in the MS complex. The constraint region computation takes $O(m)$ time in parallel -- arcs that cross or touch the boundary are identified in a first pass in $O(1)$ time followed by a neighborhood traversal in $O(m)$ time for each boundary arc. Reconnection after a critical point pair cancellation takes $O(m^2)$ time. So, the simplification takes $(nm^2/k)$ time if $O(n)$ cancellations are performed using a decomposition into $k$ subgrids.

\subsection{Implementation}
\label{sec:simplification_implementation}
The first step of parallel MS complex simplification involves creating subgrids whose size is determined based on the available CPU parallelism. We perform this in parallel, with each thread creating and storing its subgrid information independently in a shared array. 
Next, we compute the constraint region for each subgrid boundary. Using all available OpenMP threads, we mark boundary arcs and their end point nodes in parallel using a boolean array. Further, in parallel, we mark all arcs that are adjacent  to these boundary arcs. 

The final step requires the serial cancellation of each subgrid’s valid persistent arcs. We adopt previous approaches that utilize a priority queue, iteratively pop the top of the queue, check if the arc is eligible to be canceled, and then cancel it. Deleted arcs are removed in a lazy manner, new arcs are inserted into the MS complex and potentially into the priority queue. All arcs created due to a cancellation could potentially be adjacent to a boundary arc and hence belong to the constraint region. We handle this by exploring both conservative and non-conservative approaches. The former approach postpones the cancellation of all newly created arcs until the final serial cancellation step. The latter approach checks if either end point node of a newly created arc is a boundary node. If yes, the newly created arc lies within the constraint region and is not included into the priority queue. Else, it is included into the priority queue and considered for cancellation.  

We repeat this procedure iteratively for different directions and offsets. 
We experimented with two sequences for iterative slicing. The first sequence (XYZ + XYZ) performs six iterations. The initial iterations subdivide the grid along the X-axis followed by Y and Z, the subsequent three iterations subdivide the grid in the same order but with a 50\% offset. The second sequence (XX + YY + ZZ) also performs six iterations, first by subdividing along X-axis followed by another subdivision along X-axis with a 50\% offset. We investigated both sequences in combination with both conservative and non-conservative approaches described earlier.

\subsection{Discussion}
We investigated multiple variants of the subdivision scheme and methods for scheduling the cancellations. We now discuss plausible reasons for their deficient performance. Small subgrid sizes and simultaneous slicing in X, Y, and Z directions within  a single iteration resulted in poor simplification progress. We ascribe this to the potential creation of a large number of invalid boundary arcs, which in turn serialize the simplification. We tested multiple offset values for shifting subgrid boundaries and observed declining progress in simplification after successive iterations. Increasing the number of subgrids led to poor runtime performance and a significant increase in the number of strangulations, both of which indicate the presence of several boundary and high valence arcs. From our experiments, we found that a subdivision into two subgrids resulted in the best performance. We also explored the option of including an iteration where we cancel all zero persistence arcs in a serial step prior to shifting boundaries or switching slicing directions, but there was no noticeable difference in performance.

In addition to the above, we investigated a generic approach for data parallel simplification, namely by computing a maximal independent set. A maximal independent set is computed on the 1-skeleton of the MS complex using the 3-link neighborhood to establish independence between arcs. The independent set comprises the candidate set of arcs to be canceled. Canceling all valid arcs in the independent set in parallel yielded a poor rate of progress in the simplification process, especially for larger datasets. We suspect that the presence of large valence arcs inhibits a sufficiently large independent set from being identified in each iteration. Experiments indicated a large disparity between the size of the independent set and the size of the 1-skeleton as the dataset sizes grew.

\section{Computational Experiments}
\label{sec:experiments}
All steps of the computational pipeline execute on the GPU, beginning from gradient pair assignment, critical point identification, computing 1-saddle -- 2-saddle connections, and computing the saddle-extrema arcs. We perform detailed experiments to compare performance against pyms3d~\cite{shivashankar3d,mscsoftware2017,mscomplexsoftware} and \ttk~\cite{tierny2017topology}, which report the best runtimes for computing the MS complex on shared memory processors. We also perform experiments to analyze the effectiveness of the various steps of the parallel algorithm and to study the available parallelism for the different computational steps.

All experiments are performed on an Intel Xeon Gold 6130 CPU~@~2.10 GHz powered workstation with 16 cores (32 threads), 32 GB RAM, and an Nvidia GeForce GTX 1080 Ti card with 3584 CUDA cores and 11 GB RAM.
We use CUDA version 10.1 and its inbuilt Thrust and cuSPARSE libraries. We set the number of CUDA threads per block as 512 with 64 blocks in a grid, and use an identical set of parameters and best optimization flags for pyms3d and \ttk. We use the open source script available for \ttk to run the comparative experiments. We note that the methods described by Gyulassy et al.~\cite{gyulassy2012computing, gyulassy2014conforming, gyulassy2018shared} and Bhatia et al.~\cite{bhatia2018topoms} focus on improving geometric accuracy, thus yielding larger runtimes. All experiments are performed on popular volumetric simulation and imaging datasets~\cite{openSciVizWeb}. 

We also compare runtimes of our OpenMP based parallel simplification with prior work. For these experiments, we enable Intel Turbo Boost Technology to ensure optimal CPU parallel performance. Finally, an extensive collection of sanity checks and visual comparisons ensure correctness of the results.  
\begin{table*}[ht]
\centering
  \begin{center}
    \begin{tabular}{r|c|r|r|r|r||r|r|r|r|r|r}
      \textbf{Dataset} & \textbf{Size} & \textbf{Number of} & \textbf{\gmsc} & \textbf{\ttk} & \textbf{\gmsc--\ttk} &
      \textbf{pyms3d} & \multicolumn{3}{c}{\textbf{\gmsc--pyms3d Speedup}} \\
      & & \textbf{Critical Points}  & ($secs$) & ($secs$) & \textbf{Speedup} & ~\cite{shivashankar3d} ($secs$) & \textbf{Overall} & \textbf{Parallel BFS} & \textbf{Path Counting}\\ \hline
      Silicium & 98$\times$34$\times$34 & 1345 & 0.31 & 0.16 & 0.5 & 0.07 & 0.2 & 14.5 & 0.1\\
      Fuel & 64$\times$64$\times$64 & 783 & 0.33 & 0.28 & 0.9 & 0.05 & 0.1 & 2.9 & 0.1 \\
      Neghip & 64$\times$64$\times$64 & 6193 & 0.37 & 0.34 & 0.9 & 0.28 & 0.8 & 23.0 & 0.3 \\
      Tooth & 103$\times$94$\times$161 & 827973 & 1.48 & 3.08 & 2.1 & 7.88 & 5.3 & 318 & 2.8 \\
      Hydrogen & 128$\times$128$\times$128 & 26725 & 0.56 & 3.37 & 6.0 & 0.91 & 1.6 & 32.6 & 0.5 \\
      Shockwave & 64$\times$64$\times$512 & 2477 & 0.46 & 1.71 & 3.7 & 0.17 & 0.4 & 5.2 & 0.1 \\
      Lobster & 301$\times$324$\times$56 & 1201727 & 3.04 & 8.35 & 2.7 & 15.44 & 5.1 & 104.2 & 2.9 \\
      Ventricles & 256$\times$256$\times$124 & 6073455 & 8.39 & 19.66 & 2.3 & 56.07 & 6.7 & 558.3 & 3.5 \\
      Engine & 256$\times$256$\times$128 & 1541859 & 5.17 & 10.95 & 2.1 & 24.98 & 4.8 & 134.1 & 2.3 \\
      Bonsai & 256$\times$256$\times$256 & 567133 & 5.42 & 19.04 & 3.5 & 38.58 & 7.1 & 92.7 & 5.4 \\
      Aneurysm & 256$\times$256$\times$256 & 97319 & 1.21 & 31.37 & 26.0 & 7.70 & 6.4 & 66.3 & 5.2 \\
      Foot & 256$\times$256$\times$256 & 2387205 & 8.21 & 20.69 & 2.5 & 47.75 & 5.8 & 152.3 & 3.5 \\
      Turbulence & 256$\times$256$\times$256 & 1474891 & 4.92 & 17.27 & 3.5 & 42.11 & 8.6 & 361.5 & 3.5 \\
      Skull & 256$\times$256$\times$256 & 5786993 & 12.93 & 27.17 & 2.1 & 75.57 & 5.8 & 343.0 & 2.6 \\
      Angio & 416$\times$512$\times$112 & 17811553 & 22.60 & 61.44 & 2.7 & 165.88 & 7.3 & 577.7 & 3.7 \\
      Isabel-Precip & 500$\times$500$\times$100 & 1705641 & 7.56 & 25.93 & 3.4 & 37.36 & 4.9 & 86.6 & 2.0 \\
      Heptane & 302$\times$302$\times$302 & 207431 & 14.29 & 47.43 & 3.3 & 15.81 & 1.1 & 79.7 & 0.6 \\
    \end{tabular}
  \end{center}
  \caption{Runtime performance comparisons between the GPU parallel algorithm \gmsc, pyms3d~\cite{shivashankar3d,mscsoftware2017}, and \ttk~\cite{tierny2017topology}. We observe good speedup in runtime on most datasets. In comparison with pyms3d, both saddle-saddle reachability and gradient path counting algorithms contribute to the improved running times. The saddle-saddle reachability algorithm shows speedup for all datasets (2.9$\times$ to 577.7$\times$), the gradient path counting algorithm has speedup for all larger datasets (2$\times$ to 5.4$\times$).} 
  \label{tab:mscresults}
\end{table*}

\begin{figure*}[!ht]
   \centering
   \includegraphics[width=1\linewidth]{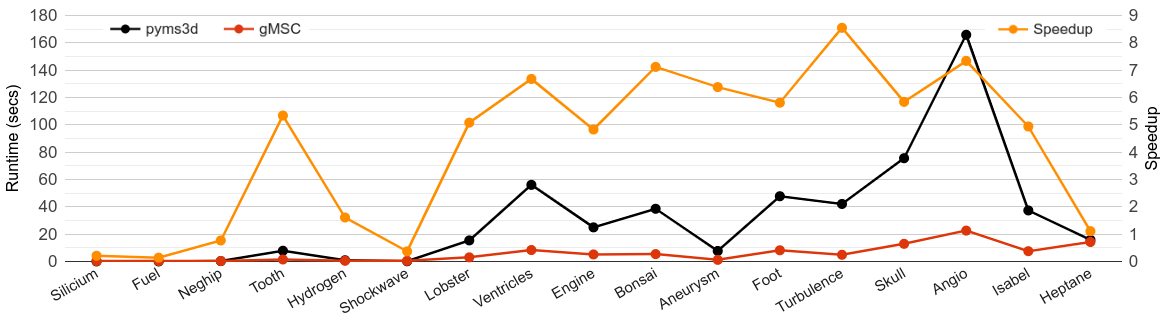}
     \caption{Runtime comparisons (left axis) between \gmsc and pyms3d~\cite{shivashankar3d,mscsoftware2017} and speedup (right axis), for increasing dataset sizes (left to right on x-axis). \gmsc outperforms pym3d for all large datasets, with significant runtime differences and speedups between 4.8$\times$ and 8.6$\times$. Runtimes for \gmsc do not increase significantly with  dataset size, unlike pyms3d, thus indicating good scalability.}
     \label{fig:runtimes_speedup}
\end{figure*}

\begin{figure*}[!ht]
   \centering
   \includegraphics[width=1\linewidth]{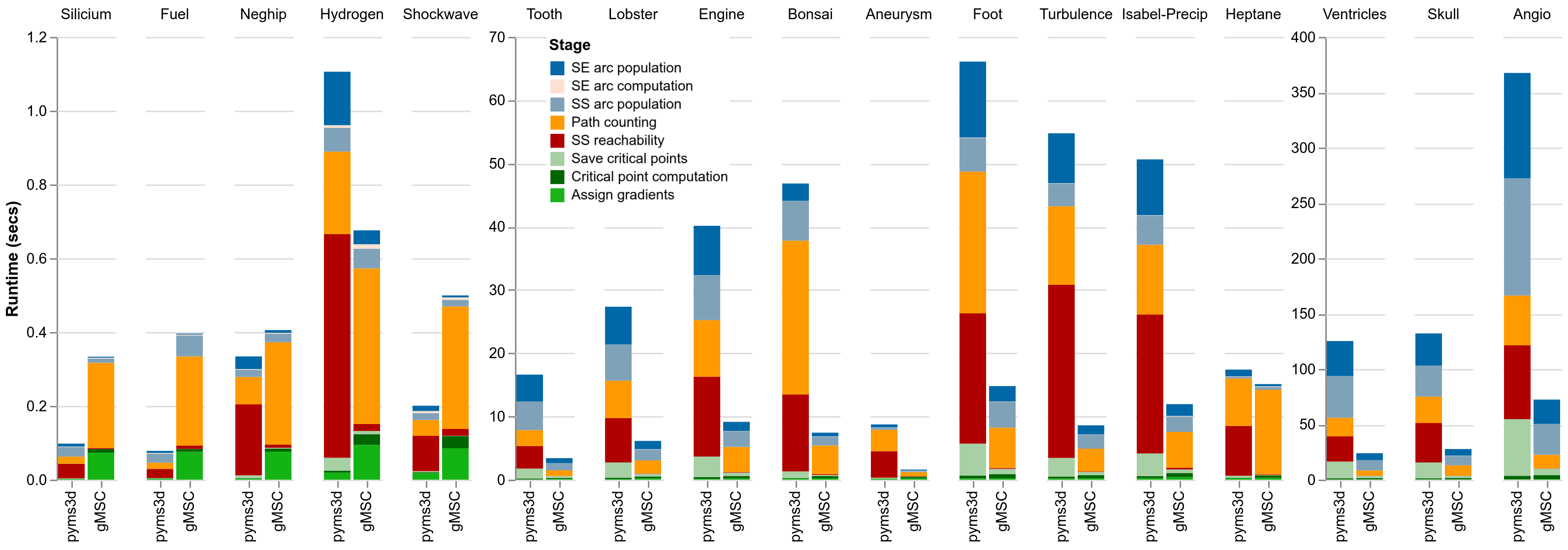}
     \caption{Runtime comparisons between \gmsc and pyms3d for each stage of the computational pipeline. Runtimes are shown in three charts to facilitate comparisons because they vary significantly between datasets. The datasets are placed in increasing order of size within each chart. We observe significant improvements to saddle-saddle (SS) reachability and path counting steps, which are the major bottlenecks in pyms3d for large datasets. \gmsc populates the data structures containing critical points, saddle-saddle (SS), and saddle-extrema (SE) arcs in parallel. All computationally expensive stages of the pyms3d pipeline are improved in \gmsc, while retaining the efficient stages.}
     \label{fig:runtimes_splitup}
\end{figure*}

\subsection{MS complex computation : runtime comparisons}
We studied the runtime performance of pyms3d and found that assigning gradient pairs and marking (computing+saving) critical points account for up to 34.4\% (Angio) of the total MS complex computation time and computing extrema paths accounts for up to 4\% time (Fuel). In contrast, marking reachable saddle-saddle pairs requires up to 68.5\% time (Neghip) due to its serial nature and the gradient path counting step requires up to 63\% (Bonsai) due to the use of coarse grained CPU parallelism, which is typically affected by load imbalance issues and excessive serialization. So, the saddle-saddle arc computation is indeed the bottleneck. We also note that a large fraction of time within gradient pair assignment and marking critical points is the step where critical points are saved. We improve the runtime of this step with the aid of data parallel primitives.

\myparagraph{Total computation time.}
\autoref{tab:mscresults} compares the runtimes with pyms3d and shows the speedup for individual stages of the pipeline. We observe 4.8$\times$ to 8.6$\times$ speedup for the overall MS complex computation for larger datasets.
The parallel saddle-saddle reachability algorithm performs better on all datasets, achieving up to 577$\times$ speedup (Angio). The speedup increases with the number of critical points as expected. The path counting algorithm also performs better for larger datasets, achieving up to 5.4$\times$ speedup (Bonsai). In Silicium, Fuel, Neghip, and Shockwave, the number of saddles and junction nodes is multiple orders of magnitude smaller than those in the larger datasets. The overheads of GPU transfer time and construction of matrices dominate the runtime resulting in poor scaling. In contrast, pyms3d processes each 2-saddle in parallel and each traversal originating at a 2-saddle terminates early given the fewer number of junction nodes. This is likely the reason for its superior performance on these  datasets. The number of junction nodes is above this threshold in other datasets and \gmsc is able to handle the explosion in the number of gradient paths more effectively than pyms3d. 
We find that \gmsc outperforms \ttk as well, for all datasets except the small ones, with a speedup between $2\times$ - $6\times$, and a high of 26$\times$ (Aneurysm). We also note that \ttk outperforms pyms3d in many instances. While \ttk cites the same algorithm as pyms3d, we believe this superior performance can be attributed to \ttk leveraging parallelism effectively in all stages of the computation, the use of cache efficient data structures, and advanced memory management techniques.

\autoref{fig:runtimes_speedup} shows the runtimes for all datasets and the speedup obtained over pyms3d. With the exception of smaller datasets such as Silicium, Fuel, Neghip, and Shockwave, \gmsc outperforms pyms3d for all other datasets, with significant runtime differences for larger datasets. We observe speedups between 4.8$\times$ to 8.6$\times$ with the exception of Heptane. We note that \gmsc runtimes do not exhibit significant increase with an increase in dataset size in contrast with pyms3d. We attribute this to pyms3d's time consuming saddle-saddle reachability step and to its coarse grained parallelism which causes an increase in the work per thread for gradient path counting as the number of critical points (saddles) increase. We also note that pyms3d performs relatively well with Isabel and Heptane despite their sizes. \gmsc encounters the largest number of sparse matrix multiplications  (730) in Isabel. The minor speedup in Heptane could be due to the significantly high path multiplicities (order of  10\textsuperscript{10}) in multiple saddle-saddle paths and the denser matrices (larger number of junction nodes and critical points) resulting in more time required for each matrix operation.

\myparagraph{Runtime for individual steps.}
\autoref{fig:runtimes_splitup} shows the contributions of the key steps of the algorithm towards the overall runtime. The saddle-saddle path computation dominates the computational runtime and is the key contributor towards the overall speedup. Runtimes for assigning gradient pairs, computing critical points, and computing extrema connections are equal because pyms3d and \gmsc implement the same algorithm with the only difference being that the former uses OpenCL and the latter uses CUDA. Further, the runtimes for these steps is a small fraction of the overall runtime. Assigning gradient pairs in \gmsc takes a larger runtime of roughly 0.1 seconds for the smaller datasets, when compared to pyms3d. We suspect this consistent, irreducible runtime to be due to the underlying data structure setup time on CUDA, which is not offset by sufficient parallel work available in these datasets. Saving critical points on the CPU is a bottleneck in pyms3d. \gmsc achieves a speedup of up to 8.5$\times$ (Angio) for this step. The stacked bar chart shows that the saddle-saddle reachability and gradient path counting steps in pyms3d are its biggest bottlenecks. \gmsc achieves significant improvements to these two steps.

We observe that the number of junction nodes is always comparable to the number of critical points, leading to sufficient parallelism in larger datasets for both saddle-saddle reachability and gradient path counting steps. Smaller datasets have fewer critical points and junction nodes which prevents them from fully exploiting the available parallelism. For example, the saddle-saddle reachability step in \gmsc outperforms pyms3d by a large factor but the gradient path counting step becomes a bottleneck for Silicium, Fuel, Neghip, Hydrogen, Shockwave, and Heptane.

The time taken to populate the MS complex data structure on the CPU is not included in the runtimes reported for computation. \autoref{fig:runtimes_splitup} shows runtimes for populating the data structures that store saddle and extrema paths. \gmsc populates the data structures in parallel on the CPU whereas pyms3d populates them serially. We note that this population is a bottleneck in many datasets, with the exceptions of datasets that are small in terms of size or number of critical points such as Silicium, Fuel, Neghip, Shockwave, Aneurysm and Heptane. Our saddle-saddle arc and saddle-extrema arc population routines achieve speedup up to 4.5$\times$ (Bonsai) and 5.4$\times$ (Engine and Turbulence). 

\begin{figure*}[!ht]
   \centering
   \includegraphics[width=1\linewidth]{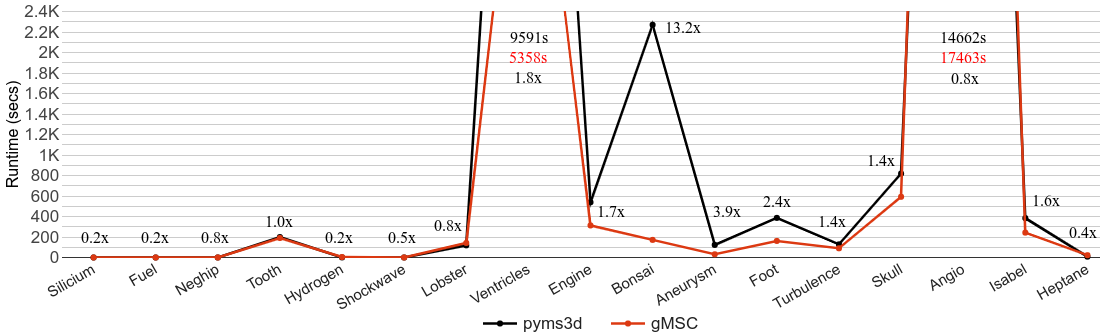}
    \caption{Runtime comparison between \gmsc and pyms3d for MS complex simplification  for increasing dataset sizes (left to right). \gmsc outperforms pym3d for all large datasets with the exception of Angio and Heptane. We observe speedups up to 13.2$\times$ (Bonsai).}
    \label{fig:simplification_runtimes_speedup}
\end{figure*}

\subsection{MS complex computation : performance analysis}
The speedups achieved by \gmsc are due to the fine-grained parallelism that it leverages at each stage of the computational pipeline. We now present results of our investigation on the available parallelism  at different steps of the computation. This investigation helps us prioritize efforts towards runtime improvement and study the efficiency of the algorithm and its implementation.

\myparagraph{Junction nodes and saddle statistics.}
The efficiency of the parallel BFS, DAG minor construction, path counting, and matrix multiplication operations relies on the number of junction nodes and saddles, which determine the available parallelism. We collect statistics on these nodes for all datasets and the results inform our choice between CPU and GPU parallelism. We provide a summary of our findings here. The detailed statistics are available in Table~1 in the supplementary material.

We observe that the number of junction nodes often exceeds the number of critical points for all dataset sizes, sometimes by an order of magnitude. There are a few exceptions (Tooth, Ventricles, and Angio) where the counts are comparable. The number of saddles and the total number of critical points are comparable, which implies that the number of extrema are orders of magnitude smaller. pyms3d executed path counting in parallel across all 2-saddles on the CPU, which led to a serialized tracing of long paths within each thread. The large number of junction nodes indicates a good fit for fine-grained parallelism on the GPU, thus motivating our design of the DAG minor construction and the GPU path counting algorithms. We also note that the number of junction nodes is sufficiently large to exploit GPU parallelism. In smaller datasets, the benefit from the parallelism is not significant enough to result in a good overall speedup. The number of junction nodes also helps quantify the extent of exponential growth in the number of paths between saddles.

\myparagraph{Matrix sparsity.}
The number of junction nodes and saddle critical points also determines the sizes of the matrices obtained from the DAG minor construction, which in turn impacts the memory utilization. The \emph{sparsity ratio} is the ratio of the number of zero valued entries in the matrix to its total size. We analyze all input and output matrix sparsity ratios. 
We study the sparsity ratios of the input matrices $A$\textsubscript{$1s-j$}, $B$\textsubscript{$j-j$}, $B$\textsuperscript{*}\textsubscript{$j-2s$}, and $D$\textsubscript{$1s-2s$} obtained from the DAG minor construction. Detailed numbers are in Figure~1 in the supplementary material.

The sparsity ratios of matrices $A$\textsubscript{$1s-j$} and $B$\textsuperscript{*}\textsubscript{$j-2s$} are nearly identical. $B$\textsubscript{$j-j$} displays the highest sparsity ratios indicating its highly sparse nature across all datasets, whereas the final output matrix displays a low sparsity ratio. 
We notice an overall increase in sparsity ratio with an increase in dataset size. Smaller datasets exhibit lower sparsity ratios, with the exception of Tooth, whose behavior may be attributed to noise (it contains a disproportionately large number of critical points and junction nodes). Among large datasets, we note highly sparse matrices with the exceptions of Aneurysm and Heptane, which are cleaner datasets and contain fewer number of critical points.

\myparagraph{Matrix multiplication statistics.}
The iterative multiplication of $A$\textsubscript{$1s-j$} and $B$\textsubscript{$j-j$} is a bottleneck in the matrix operations. The number of iterations depends on the length of the path between a source 1-saddle at level-0 and the destination junction node at the highest level in the DAG minor. The length of this path directly impacts the running time of the matrix operations, hence we study this statistic for all datasets. Detailed statistics are available in Table~1 of the supplementary material.
We observe small path lengths in smaller datasets like Silicium, Fuel, Neghip, and Tooth. Path lengths become longer for the larger datasets, resulting in a maximum of 730 iterations required for the Isabel-Precip dataset. Some exceptions include datasets such as Ventricles, Turbulence, and Angio which display shorter paths despite their size. If we consider the number of critical points as the point of comparison, Shockwave displays a large number of iterations for a smaller number of critical points. In contrast, Tooth contains a large number of critical points, but requires few iterations. Ventricles, Turbulence, and Angio display shorter paths from the perspective of critical points as well. 

\myparagraph{Memory usage.}
The size of datasets that \gmsc can handle is dictated by the available GPU memory. We performed experiments on six datasets to understand the memory footprint and limits on data size that can be handled. The package gpustat~\cite{gpustat} was used to monitor real-time usage of GPU memory, which in turn helped estimate memory consumption statistics (see Table~4  of supplementary material). The GPU memory available to \gmsc was found to be between 8--11 GB, after accounting for background processes. Experiments on all six datasets showed a 100\% memory usage. We estimated the memory consumed by noting the sizes of data structures allocated on the GPU up to the point of termination. Total memory allocated in some cases exceeds 11 GB because this count includes memory allocated for data structures that are freed after use. The biggest consumption of memory is due to arrays that store the input scalar field (up to 9\% of total memory), gradient pair information (up to 18\%), critical points and attributes (30\%--80\%), source saddle information and visited flags ($\sim$60\%). The size of these arrays are directly proportional to the size of the dataset.  One approach towards addressing this memory limitation is to partition the data into chunks, compute the MS complex for each chunk, and stitch the pieces together~\cite{shivashankar3d} but extending this approach to the GPU remains a challenge. 


\subsection{MS complex simplification : performance analysis}
\begin{figure}[!ht]
   \centering
   \includegraphics[width=1\linewidth]{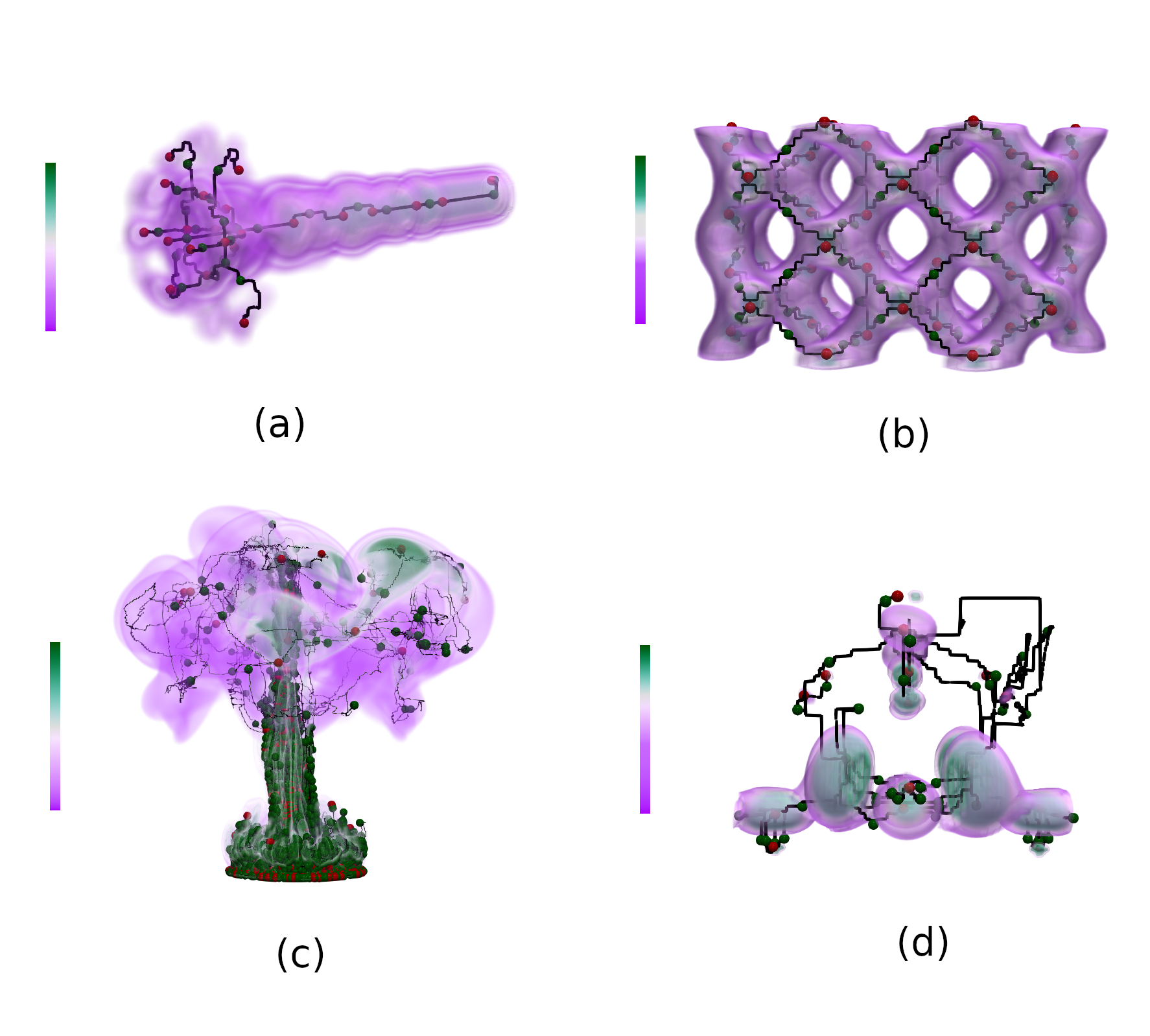}
     \caption{Saddle-maximum arcs in the simplified MS complex.  Ascending arcs between 2-saddles (green) and maxima (red) overlaid on the volume rendered field for (a) Fuel, (b) Silicium, (c) Heptane, and (d) Neghip datasets.}
     \label{fig:simplified-msc-volrendering}
\end{figure}

We quantify the performance of the parallel MS complex simplification algorithm by considering the runtimes, residual number of critical points, residual arcs, and the residual number of strangulation configurations (whose persistence is below the simplification threshold). 
We observe that our parallel algorithm outperforms its serial counterpart in pyms3d with respect to runtimes for all large datasets while achieving equivalent quality as quantified by the above-mentioned counts. \autoref{fig:simplified-msc-volrendering} shows the saddle-maximum arcs from the simplified MS complex for a few datasets.

Among the four configurations discussed in \autoref{sec:simplification_implementation}, non-conservative cancellation together with the XX + YY + ZZ  sequence of subdivision emerged as the best performer. A partition into two subgrids in each iteration yielded the best runtimes. We discuss the performance of this configuration below. Detailed analysis of the remaining three configurations are available in Figures~3 and~4 in the supplementary material. 
%

\begin{figure}[!ht]
   \centering
   \includegraphics[width=1\linewidth]{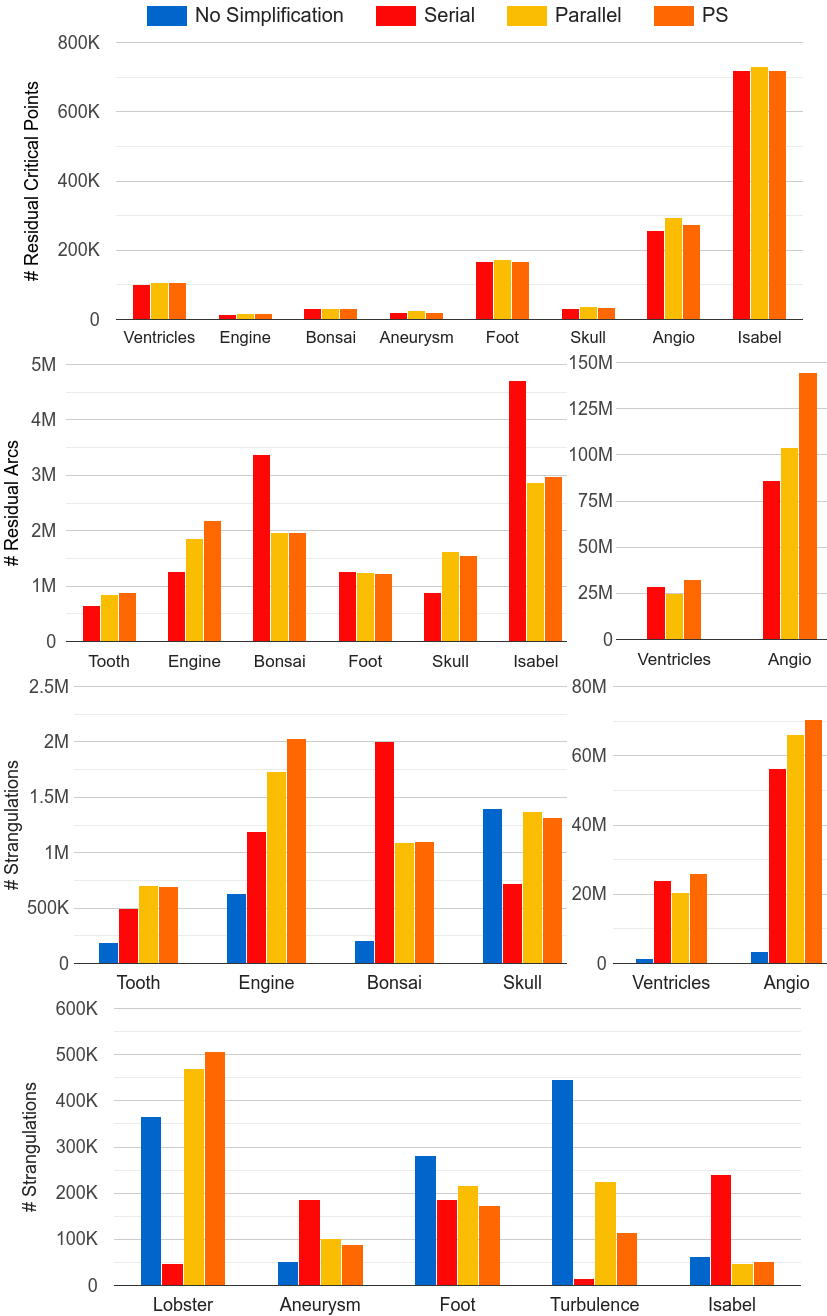}
     \caption{Plot of the residual number of critical points, arcs, and strangulations (whose persistence is below the threshold) in the simplified MS complex for both serial and parallel simplification (5\% threshold). PS denotes the execution of parallel simplification using the (XX + YY + ZZ) sequence of subdivision, followed by serial simplification. It is well established that the sequence of cancellation affects the residual numbers\cite{gyulassy2006simplification}. (top to bottom)~All large datasets show good matches for residual critical points. The number of residual arcs also match closely with the exception of Angio, which shows an increase in PS. Residual arcs in Bonsai and Isabel are fewer in PS than in serial simplification. The number of strangulations are comparatively larger after PS simplification, with a few exceptions.
     }
     \label{fig:simplification_stats}
\end{figure}

\myparagraph{Simplification time.}
\autoref{fig:simplification_runtimes_speedup} shows the runtimes for \gmsc’s parallel MS complex simplification in comparison with pyms3d’s serial simplification. Small datasets such as Silicium, Fuel, Neghip, Tooth, Hydrogen, Shockwave and Lobster show minimal or no improvements due to their small sizes enabling efficient serial simplification. We also observe a prominent increase in strangulations below the 5\% threshold in Hydrogen and Lobster for all configurations in parallel simplification. All other datasets with the exception of Angio and Heptane show significant runtime improvements, with a maximum speedup of 13.21$\times$ seen in Bonsai. We suspect Angio’s large size and structure to be contributing factors towards a large number of residual arcs and strangulations (see Figure 13), which indicate poor progress in simplification and a resulting runtime cost. Heptane is a relatively cleaner dataset with most critical points getting cancelled easily, resulting in fewer residual critical points, arcs and strangulations. In such clean datasets, serial simplification performs well. We also note that the geometric distribution of critical points could lead to load imbalance in the subgrids. For example, in Heptane, we notice a cluster of critical points in the centre, see \autoref{fig:simplified-msc-volrendering} (c), and the subgrid runtimes further indicate a serialized cancellation.

\myparagraph{Residual critical points.} 
We count the number of critical points after simplification for a quantitative comparison and perform a visual overlay for multiple persistence thresholds to determine the efficacy of the parallel simplification. 
\autoref{fig:simplification_stats} plots the residual number of critical points in the larger datasets for the serial simplification, parallel simplification across subgrids (parallel), and parallel simplification across subgrids followed by serial simplification of all critical point pairs that were not canceled (PS). All methods use a 5\% persistence threshold. We observe similar counts after serial and PS simplification. Statistics for other datasets is available in Figure~2 in the supplementary material. 
We attribute the larger number of residual critical points in Hydrogen, Lobster, and Turbulence to the  creation of several strangulations. Note that the parallel cancellation step  tends to result in a larger residual number, especially in smaller datasets such as Fuel and Neghip. This effect is neutralized after the subsequent serial cancellation.

\myparagraph{Residual arcs.}
Next, we compare the number of residual arcs in the MS complex. \autoref{fig:simplification_stats} shows a close match among all large datasets except Angio, which contains a larger number of residual arcs after parallel simplification. The number of residual arcs is smaller in Bonsai, Isabel, and Aneurysm.  We attribute the larger number of arcs in Lobster and Turbulence  to the presence of a large number of strangulations after cancellation in parallel across subgrids. We observe an increase in the number of residual arcs after parallel cancellation in all small datasets, which is neutralized by the subsequent serial cancellation. Detailed statistics are available in Figure~2 in the supplementary material.

\myparagraph{Strangulations below the persistence threshold.}
We compute and compare the number of strangulation configurations below a 5\% threshold in the simplified MS complex. This count is a measure of how difficult it is to further simplify the MS complex. As a baseline, we compute this parameter before applying the cancellation operations (no simplification). \autoref{fig:simplification_stats} shows that the two methods produce similar number of strangulations in the larger datasets. Angio, Turbulence, and Lobster contain a higher number of strangulations when simplified in parallel. The counts after parallel cancellations are mildly higher in Skull and Engine and lower in Bonsai, Aneurysm, and Isabel. The counts (PS) are similar in smaller datasets even though they are larger after parallel cancellation. Figure~3 in the supplementary material compares the number of strangulations for all configurations. The XX + YY + ZZ sequence of subdivision together with the non-conservative approach for canceling arcs in the constraint region consistently performs better.

\myparagraph{Distribution of critical points.}
We compare the distribution of different types of residual critical points, both quantitatively and visually (Figure~3 in supplementary material). Many datasets exhibit nearly identical results for the serial and parallel approaches. In some datasets (Engine, Hydrogen, Lobster, Skull, Tooth and Turbulence), we observe that the number and positions of minima and maxima match well but the number of saddles is larger after parallel simplification (PS), with a correspondingly larger number of strangulations.  We suspect that many of the strangulations are saddle-saddle connections. The presence of these strangulations adversely affects the runtimes in smaller datasets. We attribute this to the excess time spent towards the creation of these arcs and how their presence affects progress in the simplification process.

\section{Conclusions}
We have introduced a novel parallel algorithm that computes the MS complex of a 3D scalar field defined on a grid. It is the first completely GPU parallel algorithm developed for this task and results in superior performance with notable speedup. Our approach is the first of its kind to leverage data parallel primitives coupled with matrix multiplication as a means of gradient path traversal. The algorithm follows an approach to parallelism that avoids locks or synchronization.  We also introduce a data parallel algorithm for MS complex simplification which achieves superior runtimes for the relatively larger sized data in our test set while providing an improved or comparable quality of the simplified complex. Parallel approaches to MS complex simplification have not been reported earlier even though fast methods are available for topological simplification of scalar fields~\cite{lukasczyk2020localized}. Developing these methods towards improved parallel algorithms for MS complex is an interesting topic for future work.

\ifCLASSOPTIONcompsoc
  \section*{Acknowledgments}
\else
  \section*{Acknowledgment}
\fi
This work is partially supported by a Swarnajayanti Fellowship from the Department of Science and Technology, India (DST/SJF/ETA-02/2015-16), a Mindtree Chair research grant, and an IoE grant from Indian Institute of Science, Bangalore.

\bibliographystyle{abbrv}
\bibliography{references}


\begin{IEEEbiography}[{\includegraphics[width=1in,height=1.25in,clip,keepaspectratio]{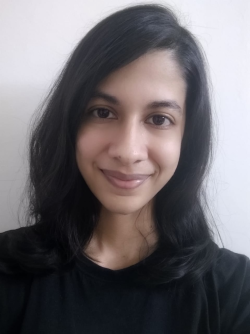}}]{Varshini Subhash} is a Project Assistant at the Visualization and Graphics Lab in the Department of Computer Science and Automation at Indian Institute of Science, Bangalore. She received her Bachelor's degree in Mechanical Engineering from Manipal Institute of Technology. Her research interests include parallel computing, algorithms and computational geometry.
\end{IEEEbiography}


\begin{IEEEbiography}[{\includegraphics[width=1in,height=1.25in,clip,keepaspectratio]{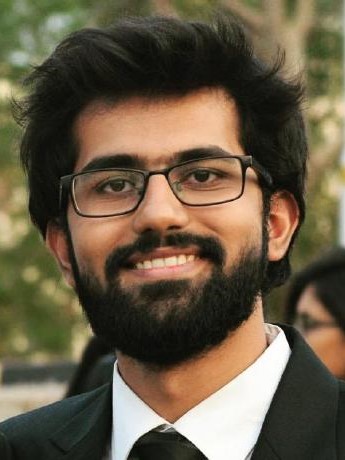}}]{Karran Pandey} is a Project Assistant at the Visualization and Graphics Lab in the Department of Computer Science and Automation at Indian Institute of Science, Bangalore. He received his M.Sc. in Mathematics and B.E. in Computer Science from BITS Pilani Hyderabad Campus. His research interests include scientific visualization, computational topology and geometry processing. 
\end{IEEEbiography}


\begin{IEEEbiography}[{\includegraphics[width=1in,height=1.25in,clip,keepaspectratio]{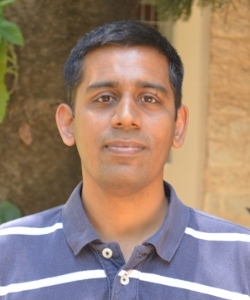}}]{Vijay Natarajan} is the Mindtree Chair Professor in the Department of Computer Science and Automation at Indian Institute of Science, Bangalore. He received the Ph.D. degree in computer science from Duke University. His research interests include scientific visualization, computational topology, and computational geometry. In current work, he is developing topological methods for time-varying and multi-field data visualization, and studying applications in biology, material science, and climate science.
\end{IEEEbiography}

\end{document}